\documentclass{article}
\usepackage{subfigure}
\usepackage{bm,latexsym,amsmath,amssymb,amsfonts,fancyhdr,color,graphicx,multirow,slashed,cite}
\usepackage[a4paper,bottom=3cm,top=2.5cm,head=0mm,width=17cm,dvipdfm]{geometry}
\usepackage[usenames,dvipsnames,svgnames,table]{xcolor}
\usepackage[colorlinks=true,
            linkcolor=blue,
            urlcolor=blue,
            citecolor=green,          
						bookmarks=true,
						bookmarksnumbered=true,
						breaklinks=true,
						pdfpagemode=Fullscreen,
						pdfstartview=FitBH]{hyperref}
\usepackage[dotinlabels]{titletoc}
\usepackage{titlesec}
\usepackage{authblk,ulem}

\numberwithin{equation}{section}
\allowdisplaybreaks[4]

\titlelabel{\thetitle.\quad \hspace{-0.8em}}
\titlecontents{section}
              [1.5em]
              {\vspace{4mm} \large \bf}
              {\contentslabel{1em}}
              {\hspace*{-1em}}
              {\titlerule*[.5pc]{.}\contentspage}
\titlecontents{subsection}
              [3.5em]
              {\vspace{2mm}}
              {\contentslabel{1.8em}}
              {\hspace*{.3em}}
              {\titlerule*[.5pc]{.}\contentspage}
\titlecontents{subsubsection}
              [5.5em]
              {\vspace{2mm}}
              {\contentslabel{2.5em}}
              {\hspace*{.3em}}
              {\titlerule*[.5pc]{.}\contentspage}

\newcommand{\titledef}{Hadron Collider Probes of the Quartic Couplings\\ of Gluons to the Photon and $Z$ Boson} 

\hypersetup{ pdfauthor = {Shao-Feng Ge},
	     pdftitle = {\titledef}, 
	     pdfsubject = {}, 
             pdfkeywords = {}, 
	     pdfcreator = {LaTeX with hyperref package},	     
	     pdfproducer = {dvips + ps2pdf} }

\newcommand{\tr}{\mbox{Tr}}

\newcommand{\bee}{\begin{equation}}
\newcommand{\ene}{\end{equation}}
\newcommand{\bea}{\begin{eqnarray}}
\newcommand{\ena}{\end{eqnarray}}

\def\fb{\, {\rm fb}}

\def\ab{\, {\rm ab}}
\def\tev{{\rm TeV}}
\def\gev{{\rm GeV}}

\def\cald{\mathcal{D}}

\def\calg{\mathcal{G}}

\def\call{\mathcal{L}}
\def\calm{\mathcal{M}}

\def\calo{\mathcal{O}}
\def\calp{\mathcal{P}}

\definecolor{gesfblack}{rgb}{0,0,0}

\definecolor{gesfblue}{rgb}{0.08,0.42,0.76}

\definecolor{gesfgreen}{rgb}{0,1,0}

\definecolor{gesfgrey}{rgb}{0.5,0.5,0.5}

\definecolor{gesflanse}{rgb}{0.00,0.50,0.50}

\definecolor{gesfpurple}{rgb}{0.47,0.19,0.42}

\definecolor{gesfred}{rgb}{1,0,0}
\newcommand{\gred}[1]{{\color{black} #1}}
\definecolor{gesfwhite}{rgb}{1,1,1}

\definecolor{gesfyellow}{rgb}{0.7,0.4,0.3}

\newcommand{\gsec}[1]{{\hypersetup{linkcolor=red}Sec.\,\ref{#1}\hypersetup{linkcolor=blue}}}

\newcommand{\geqn}[1]{\hypersetup{linkcolor=blue}Eq.\,(\ref{#1})\hypersetup{linkcolor=blue}}
\newcommand{\gfig}[1]{{\hypersetup{linkcolor=violet}Fig.\,\ref{#1}\hypersetup{linkcolor=blue}}}
\newcommand{\gtab}[1]{{\hypersetup{linkcolor=gesflanse}Table~\ref{#1}\hypersetup{linkcolor=blue}}}

\definecolor{Orange}{cmyk}{0,0.61,0.87,0}
\definecolor{JungleGreen}{cmyk}{0.99,0,0.52,0}
\definecolor{OliveGreen}{cmyk}{0.64,0,0.95,0.40}
\definecolor{Brown}{cmyk}{0,0.81,1,0.60}
\definecolor{RoyalBlue}{cmyk}{0.71,0.53,0,0.12}
\definecolor{Gray}{cmyk}{0,0,0,0.40}
\definecolor{LightPink}{cmyk}{0.0,0.25,0,0}
\definecolor{LLightPink}{cmyk}{0.0,0.10,0,0}
\definecolor{LightBlue}{cmyk}{0.25,0,0,0}
\definecolor{LightGray}{cmyk}{0,0,0,0.2}

\usepackage{graphicx}
\graphicspath{{./figs/}}

\begin{document}
\fontsize{11pt}{12pt}\selectfont

\title{
			 \vskip -30pt
       \textbf{\fontsize{19pt}{20pt}\selectfont \titledef}} 
\author[1,2,3]{{\large John Ellis} \footnote{\href{mailto:John.Ellis@cern.ch}{John.Ellis@cern.ch}}}
\author[4,5]{{\large Shao-Feng Ge} \footnote{\href{mailto:gesf@sjtu.edu.cn}{gesf@sjtu.edu.cn}}}
\author[6,7,8,4]{{\large Kai Ma} \footnote{\href{mailto:makai@ucas.ac.cn}{makai@ucas.ac.cn}}}

\affil[1]{Department of Physics, King’s College London, Strand, London WC2R 2LS, U.K.}
\affil[2]{Theoretical Physics Department, CERN, Esplanade des Particules 1, CH-1211 Geneva 23, Switzerland}
\affil[3]{National Institute of Chemical Physics \& Biophysics, R{\" a}vala 10, 10143 Tallinn, Estonia}
\affil[4]{Tsung-Dao Lee Institute \& School of Physics and Astronomy, Shanghai Jiao Tong University, Shanghai 200240, China}
\affil[5]{Key Laboratory for Particle Astrophysics and Cosmology (MOE) \& Shanghai Key Laboratory for Particle Physics and Cosmology, Shanghai Jiao Tong University, Shanghai 200240, China}
\affil[6]{School of Fundamental Physics and Mathematical Science, Hangzhou Institute for Advanced Study, UCAS, Hangzhou 310024, Zhejiang, China}
\affil[7]{International Centre for Theoretical Physics Asia-Pacific, Beijing/Hangzhou, China}
\affil[8]{Department of Physics, Shaanxi University of Technology, Hanzhong 723000, Shaanxi, China}

\maketitle
\vspace{-1.cm}
\begin{abstract}
\fontsize{10pt}{12pt}\selectfont
We explore the experimental sensitivities of measuring
the $gg \rightarrow Z \gamma$ process at the LHC
to the dimension-8 quartic couplings of gluon pairs to the $Z$ boson and photon, 
in addition to comparing them with the analogous
sensitivities in the $gg \to \gamma \gamma$ process. These processes can both receive contributions from
4 different CP-conserving dimension-8 operators with distinct Lorentz structures
that contain a pair of gluon field strengths, 
$\hat G^a_{\mu \nu}$, and a pair of electroweak SU(2) gauge field strengths, $W^i_{\mu \nu}$,
as well as 4 similar operators containing a pair of $\hat G^a_{\mu \nu}$ 
and a pair of U(1) gauge field strengths, $B_{\mu \nu}$. 
We calculate the scattering angular distributions for $gg \rightarrow Z \gamma$ and the
$Z \to \bar f f$ decay angular distributions for these 4 Lorentz structures, 
as well as the Standard Model background. We analyze the sensitivity of 
ATLAS measurements of the $Z(\to \ell^+\ell^-, \bar \nu \nu, \bar q q)\gamma$ final states 
with integrated luminosities up to 139~fb$^{-1}$ at $\sqrt{s} = 13$~TeV, 
showing that they exclude values $\lesssim 2$~TeV for the dimension-8 operator scales,
and compare the $Z \gamma$ sensitivity with that of an ATLAS measurement of the $\gamma \gamma$
final state. We present combined $Z \gamma$ and $\gamma \gamma$
constraints on the scales of dimension-8 SMEFT operators and $\gamma \gamma$
constraints on the nonlinearity scale of the Born-Infeld extension of the Standard Model.
We also estimate the sensitivities to dimension-8 operators of experiments at possible future proton-proton colliders
with centre-of-mass energies of 25, 50 and 100~TeV, and discuss possible measurements of the $Z$ spin
and angular correlations.\\
~~\\
KCL-PH-TH/2021-95, CERN-TH-2021-215\\
~~\\
\end{abstract}

{
\setlength{\parskip}{1pt}
\tableofcontents
}

\section{Introduction}
\label{sec:intro}

In the Standard Model (SM) of particle physics, the triple and quartic
gauge couplings are fixed by gauge symmetry.
Measuring these couplings can test not only whether the gauge symmetry
is realized linearly or nonlinearly in the low-energy effective theory
of the electroweak symmetry-breaking sector \cite{Brivio:2013pma} but
also provide an interesting way of looking for possible new physics
beyond the SM \cite{Green:2016trm,Rauch:2016pai}.
Hence the search for anomalous gauge couplings is one of the priority
measurements for LHC and possible future colliders. Many studies
have been made of the present and prospective experimental sensitivities to
triple gauge couplings (TGC) and quartic gauge couplings (QGC)
between electroweak SU(2)$_L \times $U(1)$_Y$ gauge bosons.
Quartic couplings between these and the SU(3)$_c$ gluons are absent
in the Standard Model (SM), but are also allowed
in the Standard Model Effective Field Theory (SMEFT) at the level of
dimension-8 operators. These have not been studied to the same extent as
dimension-6 SMEFT operators (see \cite{Ellis:2020unq, Ethier:2021bye} and
references therein), though there have recently been studies of 
dimension-8 operators that generate QGCs between photons \cite{Ellis:2017edi} and 
between photons and gluons \cite{Ellis:2018cos}, as well as those that
generate neutral TGCs \cite{Ellis:2019zex, Ellis:2020ljj}.

The dimension-8 operators that generate QGCs also arise from loop diagrams in the SM,
via extensions of the original calculations by Heisenberg and Euler \cite{Heisenberg:1936nmg, SANC},
and may be generated by the exchanges of massive
axion, dilaton or spin-2 resonances.
One particular combination of dimension-8 interactions arises in 
Born-Infeld theory
\cite{Born:1934gh,Fradkin:1985qd,Tseytlin:1999dj,Cheung:2018oki}:
\begin{equation}
\hspace{-2mm}
{\cal L}_{\rm BISM} \; = \;  \beta^2
\left[
  1
- \sqrt{1 + \sum_{\lambda = 1}^{12} \frac {F^\lambda_{\mu \nu} F^{\lambda, \mu \nu}}{2 \beta^2}
- \left( \sum_{\lambda = 1}^{12} \frac {F^\lambda_{\mu \nu} \widetilde F^{\lambda, \mu \nu}}{4 \beta^2} \! \right)^2} \;
\right] \,,
\label{BornInfeld}
\end{equation}
where the index $\lambda$ runs over the 12 generators of the SM 
SU(3)$_c \times $SU(2)$_L \times $U(1$)_Y$ gauge groups. The parameter
$\beta \equiv M^2$ is the Born-Infeld nonlinearity scale associated with
high-scale physics. The Born-Infeld extension of SM (BISM) may have deep roots in
M-theory-inspired models, where it is related to the separation between branes~\cite{Tseytlin:1999dj}.
Since all the SM gauge group factors appear in \geqn{BornInfeld}, it generates quartic
couplings of gluons to \gred{electroweak} gauge bosons (gluonic QGCs, gQGCs) 
\cite{Ellis:2018cos}.

The cleanest strategy for searching for the photonic and gluonic QGC operators
is to study processes that do not receive SM or dimension-6 SMEFT contributions. 
For example, the diphoton final state generated by light-by-light scattering,
$\gamma \gamma \to \gamma \gamma$, provides a very
clean probe of the photonic QGCs~\cite{Ellis:2017edi}. A first LHC constraint 
on light-by-light scattering was provided
by an ATLAS measurement in heavy-ion
collisions \cite{ATLAS:2017fur} (see also \cite{CMS:2018erd}).
Its rate was found to be consistent with 
the Heisenberg-Euler prediction \cite{Heisenberg:1936nmg, SANC},
allowing lower limits ${\cal O}(100)$\,GeV to be set 
on the scale of a Born-Infeld extension of QED and other possible
dimension-8 SMEFT interactions \cite{Ellis:2017edi}.
Recently, the CMS and TOTEM collaborations updated the lower limits
on the basis of a search for the exclusive production of high-mass $\gamma \gamma$
final states in $p p$ collisions at the LHC \cite{TOTEM:2021kin}.
Diphoton final states generated by gluon-gluon scattering, $gg \rightarrow \gamma \gamma$,
provide clean probes of gluonic QGCs~\cite{Ellis:2018cos}, and
the 13\,TeV data of ATLAS with 37\,fb$^{-1}$ \cite{ATLAS:2017ayi} enabled
lower limits $\gtrsim 1$\,TeV to be placed on the scales of the 
gQGC operators \cite{Ellis:2018cos}. 

In this paper we study in more detail the present and prospective future
experimental sensitivities to the gQGC operators
involving gluons and pairs of neutral electroweak bosons, $Z$ and $\gamma$.
We present a first analysis of $gg \to Z \gamma$ scattering 
and compare this with its $gg \to \gamma \gamma$ counterpart. We
analyze ATLAS measurements of $Z \gamma$ production followed by
$Z \to \ell^+ \ell^-$ \cite{ATLAS:2019gey},
$\bar \nu \nu$ \cite{ATLAS:2018nci} and $\bar q q$ \cite{ATLAS:2018sxj}
decays, with up to 139\,fb$^{-1}$ of luminosity at $\sqrt{s} = 13$\,TeV.
The scale of the dimension-8 gQGC operators can be constrained up to
$\gtrsim 2$\,TeV, higher than obtained in the updated analysis of the diphoton channel that we also
present in the current paper. However, we also show that whereas
measurements of the $Z \gamma$ final state do not constrain the BISM
scale, the $\gamma \gamma$ process constrains its scale to $\gtrsim 5$~TeV.
Interestingly, we note that a combination of measurements of
the $Z \gamma$ and $\gamma \gamma$ final states could in principle disentangle the
contribution of different operators. Finally we display the combined
sensitivities of the $Z \gamma$ and $\gamma \gamma$ channels,
and show how the sensitivities
to dimension-8 operators could be increased in the future
by measurements at possible proton-proton colliders with centre-of-mass
energies of 25, 50 and 100\,TeV.

{The structure of the paper is as follows. \gsec{sec:gQGC:operators}
reviews the dimension-8 gQGC operators, especially those relevant to
the associated production of $Z$ and $\gamma$, as well as $\gamma \gamma$
pair-production. Then \gsec{sec:gQGC:Property} studies the
kinematics and differential cross sections of the $gg \rightarrow Z \gamma$
process at the LHC. In particular, we show how the unitary constraint
can be implemented consistently.
Following these preparations, we investigate
the experimental constraints on the gQGC operators at the LHC in
\gsec{sec:gQGC:Sens} and future colliders in \gsec{sec:future}.
Then, in \gsec{sec:improvements} we present the helicity
amplitudes for the partonic process $gg \to Z\gamma$ and analyze the
corresponding angular distributions as well as spin correlations
in $Z$ boson decay, and discuss 
the possible interest of measurements of the $Z$ spin and angular correlations. 
Our conclusions can be found in \gsec{sec:conclusion}.}


\section{Dimension-8 Gluonic QGC Operators}
\label{sec:gQGC:operators}

Although there are many possibilities for new physics at higher scales
above the EW one, the low energy effective field theory (EFT) should be
subject to the SM SU(3)$_c \times $SU(2)$_L \times $U(1)$_Y$ gauge symmetries.
A convenient way to take these symmetries into account is the SMEFT
\cite{Buchmuller:1985jz}, which includes systematically all the allowed
interactions with mass dimension $d > 4$. The extra dimensions are
compensated by inverse powers of a mass scale $M$ that is associated with
heavy new physics. The gQGC operators appear at dimension-8 level with
$1/M^4$ suppression \cite{Ellis:2018cos},
%
%
\begin{subequations}
\label{basis}
\begin{eqnarray}
  \mathcal O_{gT,0}
& \equiv &
  \frac 1 {16 M^4_0}
       \sum_a G^a_{\mu \nu} G^{a, \mu \nu}
\times \sum_i W^i_{\alpha \beta} W^{i, \alpha \beta} \,,
\\
  \mathcal O_{gT,1}
& \equiv &
  \frac 1 {16 M^4_1}
       \sum_a G^a_{\alpha \nu} G^{a, \mu \beta}
\times \sum_i W^i_{\mu \beta} W^{i, \alpha \nu} \,,
\\
  \mathcal O_{gT,2}
& \equiv &
  \frac 1 {16 M^4_2}
       \sum_a G^a_{\alpha \mu} G^{a, \mu \beta}
\times \sum_i W^i_{\nu \beta} W^{i, \alpha \nu} \,,
\\
  \mathcal O_{gT,3}
& \equiv &
  \frac 1 {16 M^4_3}
       \sum_a G^a_{\alpha \mu} G^a_{\beta \nu}
\times \sum_i W^{i, \mu \beta} W^{i, \nu \alpha} \,,
\\
  \mathcal O_{gT,4}
& \equiv &
  \frac 1 {16 M^4_4}
       \sum_a G^a_{\mu \nu} G^{a, \mu \nu}
\times B_{\alpha \beta} B^{\alpha \beta} \,,
\\
  \mathcal O_{gT,5}
& \equiv &
  \frac 1 {16 M^4_5}
       \sum_a G^a_{\alpha \nu} G^{a, \mu \beta}
\times B_{\mu \beta} B^{\alpha \nu} \,,
\\
  \mathcal O_{gT,6}
& \equiv &
  \frac 1 {16 M^4_6}
       \sum_a G^a_{\alpha \mu} G^{a, \mu \beta}
\times B_{\nu \beta} B^{\alpha \nu} \,,
\\
  \mathcal O_{gT,7}
& \equiv &
  \frac 1 {16 M^4_7}
       \sum_a G^a_{\alpha \mu} G^a_{\beta \nu}
\times B^{\mu \beta} B^{\nu \alpha} \,.
\end{eqnarray}
\end{subequations}
In order to respect the SM gauge symmetries, the gauge bosons should appear
via their field strengths, such as $G^a_{\mu \nu}$ for gluons
together with $W^i_{\mu \nu}$ and $B_{\mu \nu}$ for EW gauge bosons.
Since gluons carry QCD color, denoted by the $a$ superscript of
$G^a_{\mu \nu}$, the gQGC operators must contain an even number of gluon
field strengths, such as $G^a_{\mu \nu} G^{a, \alpha \beta}$, so as to be
colorless. The same thing applies for the SU(2)$_L \times $U(1)$_Y$
gauge boson field strengths, for example $W^i_{\mu \nu} W^{i, \alpha \beta}$
and $B_{\mu \nu} B^{\alpha \beta}$. Another symmetry to be imposed is
Lorentz invariance. There are four different Lorentz-invariant contractions,
as shown above. Hence we must consider eight gQGC operators in total.

The total and differential cross sections are largely determined by
the Lorentz structure of the gQGC operators. The
eight operators can be classified into four pairs
\{$\mathcal O_{gT,(0,4)}$,
$\mathcal O_{gT,(1,5)}$, 
$\mathcal O_{gT,(2,6)}$, 
$\mathcal O_{gT,(3,7)}$\},
each with a same Lorentz structure. The $W^i$ and $B$ fields do not correspond to
the physical mass eigenstates, which are the photon $A$ and
$Z$ boson fields, $W^3 = c_w Z + s_w A$ and $B = c_w A - s_w Z$
where $(c_w, s_w) \equiv (\cos \theta_W, \sin \theta_W)$ are the cosine and
sine functions of the weak mixing angle $\theta_W$. Each pair
$W^i W^i$ and $BB$ contains $AA$, $AZ$ and $ZZ$ contributions with coefficients
determined by $\theta_W$:
\begin{subequations}
\label{WWBB}
\begin{eqnarray}
  \sum_i W^i W^i
& = &
  W^3 W^3 + \cdots
=
  s^2_w A A
+ 2 c_w s_w Z A
+ c^2_w Z Z
+ \cdots \,,
\\
  BB
& = &
  c^2_w A A
- 2 c_w s_w Z A
+ s^2_w Z Z
\,.
\end{eqnarray}
\end{subequations}
We see that their contributions to
${gg \rightarrow Z \gamma}$ scattering via the $ZA$ combination
differ only by a sign. Consequently, they produce exactly the
same event distributions, and therefore yield identical cross sections
for common values of the dimension-8 cut-off scale. 

Only 4 independent operators contribute to $gg \rightarrow Z \gamma$.
On the other hand, in the case of ${gg \rightarrow \gamma \gamma}$,
one needs to study separately the 8 operators, as done in
\cite{Ellis:2018cos}, and the contributions of $BB$ operators
are larger than those of their $\sum_i W^i W^i$ counterparts by a
factor $\cot^4 \theta_W$. Thus the $\gamma \gamma$ and $Z\gamma$ channels
in gluon-gluon scattering are complementary, 
and their combination can help to disentangle the $W^i W^i$ and
$BB$ components within each pair.

For comparison, the Born-Infeld extension of the SM shown in \geqn{BornInfeld}
contains only one linear combination of the eight gQGC operators,
\begin{eqnarray}
  \mathcal O_{\rm BISM}
\ni
  \frac 1 {16 M^4_{\gred{BI}}}
\left[
  \tr(GG) \tr(WW)
+ \tr(G \widetilde G) \tr(W \widetilde W)
\right] \; + \; \{W \to B\} \, ,
\label{BIdim8}
\end{eqnarray}
where the traces are over the Lorentz and SM gauge group indices.
The first term involves only CP-even pairs $GG$ and $WW$ 
of gluon and EW field strengths while the second is composed of the
CP-odd combinations $G \widetilde G$ and $W \widetilde W$.
The combinations of CP-odd operators can be rewritten
in the form of \geqn{basis}, as follows:
\begin{eqnarray}
  \tr(G \widetilde G) \tr(W \widetilde W)
=
  4 \tr(G W G W)
- 2 \tr(G W) \tr(G W),
\end{eqnarray}
and similarly for $\tr(G \widetilde G) \tr(B \widetilde B)$. 
Hence the BISM contribution to gQGC is simply a linear
combination of six gQGC operators in the basis considered above, namely
\begin{eqnarray}
  \mathcal O_{\rm BISM} \; 
\ni \;
    \mathcal O_{gT,0}
- 2 \mathcal O_{gT,1}
+ 4 \mathcal O_{gT,3}
+   \mathcal O_{gT,4}
- 2 \mathcal O_{gT,5}
+ 4 \mathcal O_{gT,7}.
\label{LBISM}
\end{eqnarray}
Most importantly, these six operators share a common cut-off scale $M$
and their coefficients are fully correlated.

It is important to observe that, since the different gauge sectors appear
with equal coefficients in the Born-Infeld lagrangian \geqn{BornInfeld}, the
dimension-8 couplings of the EW gauge bosons to gluons take the diagonal form
$\sum_i W^i W^i + B B$. Hence, because of the opposite signs and equal magnitudes
of the $ZA$ terms in \geqn{WWBB}, the BI extension of the SM does not contribute
to $Z \gamma$ production, but only to $\gamma \gamma$ and $ZZ$ production.
As we shall see in \gsec{sec:gQGC:Sens}, the sensitivity to the scale of 
the BI extension of the SM at the LHC via $gg \rightarrow \gamma \gamma$ scattering
is considerably greater than that obtained from the study
of $\gamma \gamma \rightarrow \gamma \gamma$ scattering.

\section{Properties of the $gg \rightarrow Z \gamma$ and $\gamma \gamma$ Cross Sections}
\label{sec:gQGC:Property}

As discussed in \gsec{sec:gQGC:operators}, dimension-8 contributions to gluon-gluon collisions can yield $\gamma \gamma$ and $Z \gamma$ final
states. Both can provide a clean signal for probing gQGC operators,
whilst the BISM can be probed only by the $\gamma \gamma$ channel.
This section summarizes the properties of the total and differential
$gg \rightarrow Z \gamma$ and $\gamma \gamma$ cross sections, discussing the
basic features and laying the basis for the analysis of the various $Z$
decay modes in \gsec{sec:gQGC:Sens}.

\subsection{Total Cross Section and Energy Dependence}
\label{sec:xsec}

\begin{figure}[t]
\centering
\includegraphics[width=0.48\textwidth]{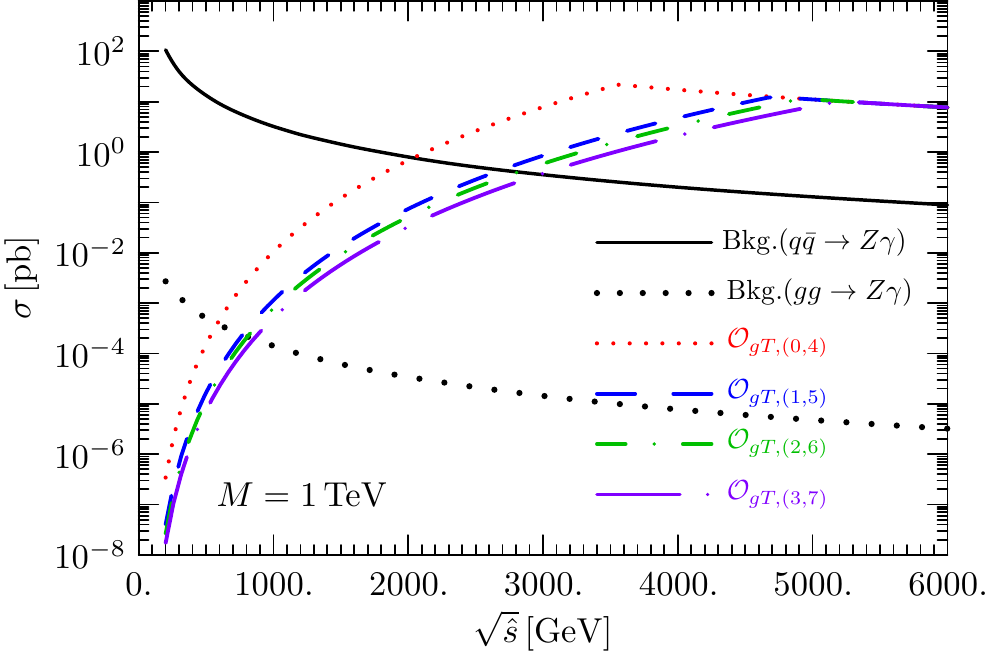}
\quad
\includegraphics[width=0.48\textwidth]{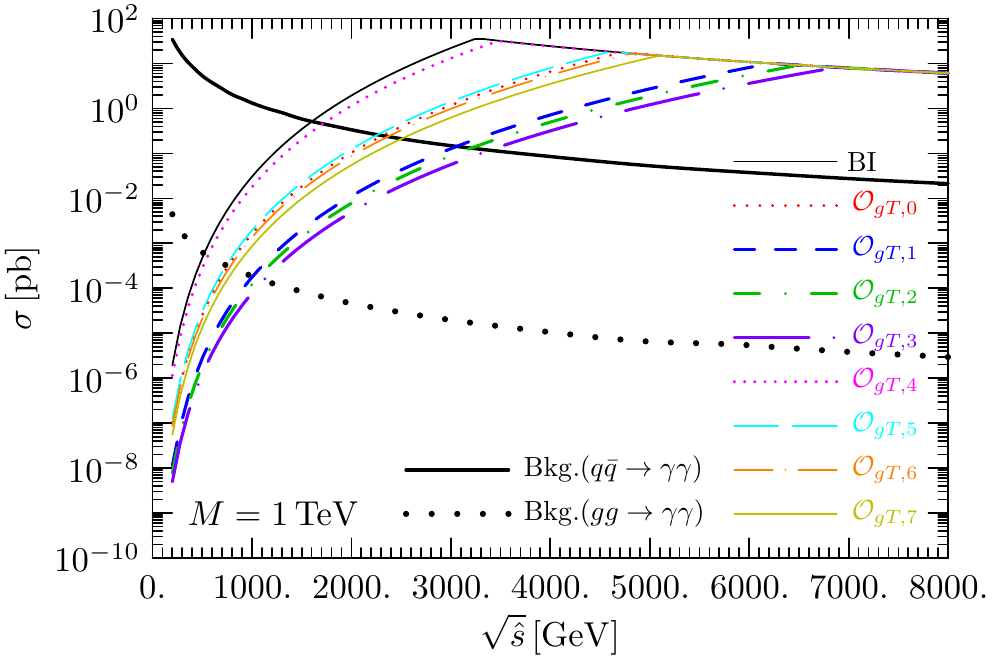}
\caption{\it The total cross sections for
$gg \rightarrow Z \gamma$ (left) and
$gg \rightarrow \gamma \gamma$ (right) generated by the
different dimension-8 gQGC operators at the parton level (colorful non-solid lines).
For illustration, we assume that the scattering amplitudes have
equal cut-off scales 
$M = 1$\,TeV until their respective
cross sections reach their unitarity limits,
which we assume to be saturated at higher gluon-gluon
centre-of-mass energies $\sqrt{\hat s}$. The cross sections
for the SM background from $q \bar q$ annihilation are shown as solid black lines.
}
\label{fig:xsec} 
\end{figure}

When regarded as a probe of new physics beyond the SM, the sensitivity of the search for a
gQGC signal is limited by the possible event rates. A conservative search strategy starts by using the
total cross section. Since each gauge field strength ($G^a_{\mu \nu}$,
$W^i_{\mu \nu}$, and $B_{\mu \nu}$) contains one derivative and hence
contributes one factor of momentum $p$ to the scattering process, the four field
strengths in a dimension-8 operator yield a scattering amplitude
\gred{$\mathcal M$} with a $p^4$
dependence. Taking into account the flux prefactor that is $\propto 1/\hat s$ for high-energy scattering, 
we find that the cross sections given by the gQGC operators grow
as $\hat s^3 \sim p^6$. Hence the signal cross sections generated by
the $\mathcal O_{gT,i}$ operators are rapidly-rising functions of the center-of-mass
energy $\hat s$ of the gluon pair:
\begin{eqnarray}
\label{eq:XS:ZA}
 \sigma_{Z \gamma, i}
=
\frac{ s_w c_w \hat{s}^{3} ( 1 - x_{Z} )^{3} }{ 2048 \, \pi \, M_{i}^8 }  
\times
\begin{cases}
1 & i = 0, 4, \\[1mm]
\dfrac{1}{120}\Big[ 13 + 4 x_{Z} + 3 x_{Z}^{2}\Big]
& i = 1, 5, \\[2mm]
\dfrac{1}{480}\Big[ 36 + 3 x_{Z} + x_{Z}^{2}\Big]
& i = 2, 6, \\[2mm]
\dfrac{1}{480}\Big[ 23 + 4 x_{Z} + 3x_{Z}^{2}\Big]
& i = 3, 7,
\end{cases}
\end{eqnarray}
where $x_Z \equiv m^2_Z / \hat s$. This effect enhances the
signal event rate significantly at higher energy, as shown in \gfig{fig:xsec},
where we see that the signal grows very fast whereas the SM background
decreases $\propto 1/\sqrt{\hat s}$. Thus one expects to find greater
sensitivities at higher-energy colliders, as we discuss later.

The steadily growing cross sections finally violate the unitarity
bound at some center-of-mass energy scale~\cite{Ellis:2018cos}.
This problem can be addressed by introducing momentum-dependent 
couplings or unitarization procedures \cite{Garcia-Garcia:2019oig}, 
or by estimating the maximum center-of-mass energy allowed by unitarity
\cite{Arnold:2008rz}. Following the procedure adopted in 
\cite{Ellis:2017edi} and \cite{Ellis:2018cos}, here we respect the
unitarity constraint by assuming that the cross section falls as
$\hat{s}^{-1}$ once the ``tree unitarity" requirement \cite{Cornwall:1974km}
is satisfied, after which the energy
scaling is the same as that of the major irreducible background due to $\bar q q \to \gamma \gamma$.
We find that this unitarity bound $\sigma \sim 1/m_{Z\gamma}^2$ is saturated at
the gluon-gluon center-of-mass energies
\bee
\sqrt{\hat s_i} 
= 
M_i \left[ \frac{s_wc_w}{2048 \, \pi} 
\left(1,\, \frac{13}{120},\, \frac{36}{480},\, \frac{23}{480} \right)
\right]^{-1/8} \,,
\ene
for $i = (0, 4), (1, 5), (2, 6), (3, 7)$, in the limit $x_Z \to 0$. For $\hat s = 1\,\mbox{TeV}^2$, the
ratio $x_Z \lesssim 0.01$ is generally negligible.

\gred{The sensitivity to the scale of new physics is quite stable with respect to
the choice of unitarization scheme. For example,  an aggressive
unitarization scheme that cuts off the signal events above the
saturation point would reduce the total event rate by a factor
of about 2. But the dependence on the new physics scale $M_i$ is
eighth power, $\sigma \propto 1 / M^8_i$, so the sensitivity is reduced by only
10\% at most. In following discussions, the same
unitarization scheme $\sigma \sim 1 / \hat s$ is applied for both
the current LHC-13\,TeV with real data and the future experiments
with projected pseudo-data.}

The total cross sections for $g g \rightarrow Z \gamma$ and $\gamma \gamma$
are summarized in \gfig{fig:xsec} as functions of
$\sqrt{\hat s}$, \gred{with $M_i = 1$\,TeV for illustration}.
The signal cross sections are much larger
than the SM background for $\sqrt{\hat s} \gtrsim 2$\,TeV,
and the tree unitarity limits are reached when
$\gred{\sqrt{\hat s_i}} = (3.58, 4.72, 4.94, 5.22) \gred{M_i}$ for $i=(0,4), (1,5), (2,6), (3,7)$ 
for the $gg\to Z\gamma$ process; 
$\gred{\sqrt{\hat s_i}} = (4.71, 6.21, 6.51, 6.88, 3.49, 4.60, 4.82, 5.10)  \gred{M_i}$ 
for $i=0, \ldots, 7$ for the $gg\to \gamma\gamma$ process;
and $\gred{\sqrt{\hat s_i} = 3.26 M_i}$ for the BI model. Even when
the unitarity bound is imposed for $g g \rightarrow Z \gamma$, replacing its rapidly-growing
cross section with the SM $1/\hat s$ scaling, the signal is almost
two orders above the SM background, as seen in the left panel of \gfig{fig:xsec}. 
For comparison, in the case of the cross section for the $\gamma \gamma$
channel shown in the right panel, the difference between
signal and background is close to three orders of magnitude. This indicates
that the $gg \rightarrow Z \gamma$
and $\gamma \gamma$ signals should be readily detectable at high-energy
hadron colliders if gQGC operators appear at the TeV scale.
Details of a more complete study are given below.

\gred{\gfig{fig:xsec} also shows the next-to-leading order (NLO) SM background due to
$gg \rightarrow Z \gamma$ via box diagram. Its cross section has been
calculated using {\tt MadGraph@NLO} \cite{Frederix:2018nkq} to calculate
the loop diagram automatically. Due to the loop factor,
this $gg$ background is suppressed by roughly four orders
of magnitude compared to its $q \bar q$ counterpart. 
In addition to its own contribution,
this loop-induced $gg \rightarrow Z \gamma$ scattering amplitude
can also interfere with the dimension-8 operator amplitude.
However, for this interference term to be comparable with the
$q \bar q$ background, the dimension-8 operator amplitude would need
to be two orders larger than the $q \bar q$ one. In this case,
the signal is much stronger than background. In the case 
that the signal is comparable with the background, i.e., the dimension-8
operator contribution is comparable with the $q \bar q$ one, the
interference term is two orders smaller. In either case, the
loop-induced SM $gg$ background is negligible. Hence, in the following
discussion we focus on the SM $q \bar q$ background and the
$gg \rightarrow Z \gamma$ signal.

We note that dimension-6 operators could in principle contribute to the same
final state $pp \rightarrow Z \gamma$.
It could well be that such operators are absent, but their presence would not invalidate our analysis. We recall that
the amplitude for a $2 \rightarrow 2$ process is dimensionless. With a coefficient that is $\propto 1/\Lambda^2$ , where
$\Lambda$ is the corresponding new-physics scale, the contribution to the amplitude should
scale as $s / \Lambda^2$ due to dimensional analysis. Comparing
with the dimension-8 contribution that scales with $s^2 / \Lambda^4$, the dimension-6
operator is suppressed in the high-energy signal region $\sqrt{\hat s} \gtrsim 1$\,TeV.
Hence dimension-6 operators are not a problem for our study of the collider sensitivities to
dimension-8 gQGC operators in this paper. We note that this argument
based on dimensional analysis applies to any dimension-6 operator, 
and is independent of its concrete form.
}

\subsection{Angular Distribution and Background Suppression}

The SM background mainly comes from $t$-channel
$q \bar q \rightarrow Z \gamma$ scattering. This process is
intrinsically suppressed at large angles by the intermediator quark propagator,
which is $\propto 1 / \hat t$ in the massless limit. For a fixed ratio
between the Mandelstam variables $\hat s$ and $\hat t$, the leading-order cross section for the SM background
is given by
\begin{equation}
  \sigma_{q\bar{q}\to Z\gamma}
=
\frac{ Q^{4}_{q} g_{E}^{4} }{ 24\pi \hat{s} }
\left[ \frac{ (T_{3}^{q})^{2} }{ Q_{q}^{2} s_{w}^{2} c_{w}^{2} } - \frac{ 2T_{3}^{q} }{ Q_{q} c_{w}^{2} } + 2\frac{s_{w}^{2}}{c_{w}^{2}} \right]
\left[  \frac{ 2(1 + x_{Z}^{2}) }{  1 - x_{Z} }\big[\tanh^{-1}{(1-\delta)} - 1 +\delta \big]
- \frac{1}{3}(1 - x_{Z}) \right],
\end{equation}
where $\delta\sim 0$ is an infrared regulator that scales as $1/\hat s$ in the high-energy limit. 
The ratio between the quark mass $m_{q}$ and $m_{Z\gamma}$,
$m_{q}/m_{Z\gamma}$, would in principle provide a natural cut-off value for the scattering angle, 
but in practice $m_{q}/m_{Z\gamma}$ is much smaller than typical kinematic cuts for event selection, 
and hence is irrelevant for our purposes.
Although the SM background decreases as $1/\hat s$, it will still dominate
the event rate if the cut-off scales $M_i$ of the gQGC operators
are much higher than the 1\,TeV used in \gfig{fig:xsec}. For this reason, it is desirable
to find kinematic cuts to suppress the SM background.

\begin{figure}[t]
\begin{center}
\includegraphics[width=0.48\textwidth]{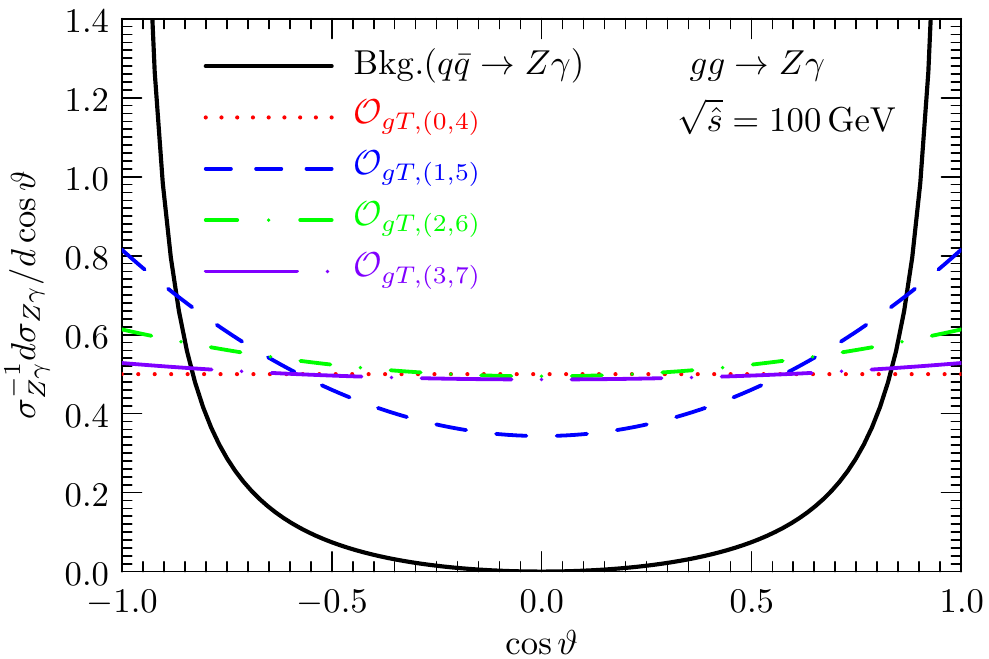}
\quad
\includegraphics[width=0.48\textwidth]{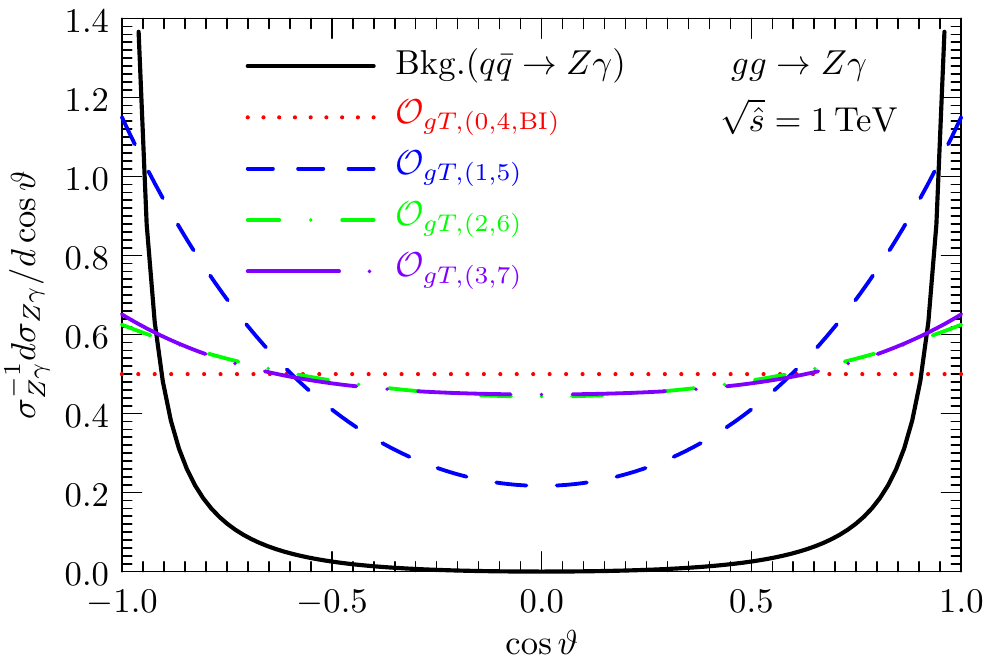}
\caption{\it
Distributions in the normalized scattering angle $\cos \vartheta$  for
$gg \to Z \gamma$ scattering from the gQGC operators (colorful
non-solid curves) and the SM background
$q \bar q \rightarrow Z \gamma$ (black solid curve).
The energy dependences are seen by comparing the left
($\sqrt{\hat{s}} = 100\,{\rm GeV}$) and
right ($\sqrt{\hat{s}} = 1\,{\rm TeV}$) panels.
}
\label{fig:dxsec}
\end{center}
\end{figure}

One such cut-off is provided by the scattering angle in the center-of-mass
frame \cite{Ellis:2018cos}, since the $Z$ and photon are emitted
as initial state radiation in the $t$-channel exchange process and hence
are typically quite forward. The differential
cross section $d \sigma / d \cos \vartheta$ \cite{Ohnemus:1992jn} is
\begin{eqnarray}
\frac {d \sigma_{q\bar{q}\to Z\gamma}}{d \cos \vartheta}
=
\frac{ Q^{4}_{q} e^4 }{ 24\pi \hat{s} }
\left[ \frac{ (T_{3}^{q})^{2} }{ Q_{q}^{2} s_{w}^{2} c_{w}^{2} } - \frac{ 2T_{3}^{q} }{ Q_{q} c_{w}^{2} } + 2\frac{s_{w}^{2}}{c_{w}^{2}} \right]
\left[  \frac{ 1 + x_{Z}^{2} }{  1 - x_{Z} }  - \frac{1}{2}(1 - x_{Z}) \sin^{2} \vartheta\right]
 \cot^{2}\vartheta,
\end{eqnarray}
where $T^q_3$ and $Q_q e$ are the quark isospin and electric charges.
As shown by the black curve in \gfig{fig:dxsec}, the SM background indeed peaks
in the forward and backward directions, where $\cos \vartheta = \pm 1$.
The comparison between the left panel ($\sqrt{\hat s} = 100$\,GeV) and
the right panel ($\sqrt{\hat s} = 1$\,TeV) shows that higher energy
leads to more forward scattering. This property applies to both
the $Z \gamma$ and $\gamma \gamma$ channels.

For comparison, the signal distribution is much less forward-backward peaked than the
SM background. The differential cross sections for the
$gg \rightarrow Z \gamma$ process generated by the dimension-8 operators are
\begin{eqnarray}
\label{eq:XS:Diff}
  \frac {d \sigma_{gg \rightarrow Z \gamma}}{d \cos \vartheta}
=
\dfrac{ s_w c_w ( 1 - x_{Z} )^{3}\, \hat{s}^{3}  }{  4096\,\pi M_{i}^{8}  }
\times
\begin{cases}
%
1
& i = 0, 4 \, ,\\[5mm]
%
\begin{aligned}
\dfrac{ 1 }{ 512 }
\Big[ & 73 + 52 \cos (2 \vartheta ) +3 \cos (4 \vartheta )  + 24 x_Z^{2} \sin^4(\vartheta )
\\&
+8 x_Z \sin^2\vartheta \big( 5 + 3\cos (2 \vartheta ) \big)
\Big] 
\end{aligned}
& i = 1, 5  \, ,\\[5mm]
%
\begin{aligned}
\dfrac{ 1 }{ 2048 }
\Big[ & 163 + 28 \cos (2 \vartheta ) + \cos (4 \vartheta )  + 8 x_Z^{2} \sin^4\vartheta
\\&
+ 8 x_Z \sin^2\vartheta \big( 3+\cos (2 \vartheta ) \big)
\Big]  
\end{aligned}
& i = 2, 6  \, , \\[5mm]
%
\begin{aligned}
\dfrac{ 1 }{ 2048  }
\Big[ & 105 + 20 \cos (2 \vartheta ) + 3\cos (4 \vartheta )  + 24 x_Z^{2} \sin^4\vartheta
\\&
+ 8 x_Z \sin^2\vartheta \big( 5+3\cos (2 \vartheta ) \big)
\Big] 
\end{aligned} & i = 3, 7
  \, .
\end{cases}
\end{eqnarray}
These angular distributions are the same as for the $gg\to \gamma\gamma$
case \cite{Ellis:2018cos} in the limit $x_Z \to 0$, reflecting the fact the 
limit $x_Z \to 0$ corresponds to a
massless $Z$ boson that is no different from a massless photon.
Since the $Z$ boson mass is significantly smaller than the typical
invariant mass $m_{Z \gamma}$ at hadron colliders,
the correction due to a finite $x_Z$ is generally negligible.

The differences between the gQGC signals (colorful non-solid lines)
and the SM background (black solid line) in \gfig{fig:dxsec} are prominent.
Since the polarizations of the incoming gluons are unknown, 
the process is symmetric under the interchange
$g( \vec{p}_{1}) \leftrightarrow g( \vec{p}_{2})$. This means that there is a
symmetry in the polar angle under $\vartheta \to \pi - \vartheta$,
under which $\sin\vartheta \to \sin\vartheta$
and $\cos\vartheta \to -\cos\vartheta$.
\gfig{fig:dxsec} shows that $\mathcal O_{gT, (0,4)}$
yield isotropic distributions, whereas the other gQGC
operators yield distributions that have small forward-backward peaks.
The four Lorentz structures have quite different distributions in general,
but we note the following similarities. At the low energy
$\sqrt{\hat s} = 100$\,GeV in the left panel,
$\mathcal O_{gT, (0,4)}$ and $\mathcal O_{gT, (3,7)}$
yield similar angular distributions, whereas in the right panel at 
$\sqrt{\hat s} = 1$\,TeV  there is greater similarity between the angular
distributions for $\mathcal O_{gT, (2,6)}$ and $\mathcal O_{gT, (3,7)}$.

Since the gQGC signals have only mild anisotropies, whereas the SM background
is concentrated in the forward and backward regions, a simple cut
on the $Z$/$\gamma$ polar scattering angle in the rest frame of the $gg$
system can suppress significantly the SM background and thereby enhance the
signal sensitivity. This feature is used in our study of the experimental
sensitivities at the LHC and future colliders in \gsec{sec:gQGC:Sens}.
The potential for distinguishing different gQGC operators via the angular distributions
is discussed further in \gsec{sec:improvements}.

\section{Search Strategy and Sensitivities at Hadron Colliders}
\label{sec:gQGC:Sens}

As discussed above, probes of the gQGC operators at hadron colliders with the processes $gg \to Z \gamma$
and $\gamma \gamma$ are potentially interesting. The signals would
dominate over the SM background when the $gg$ center-of-mass energy
reaches the TeV scale, and a simple cut on the scattering angle can further
enhance the sensitivity significantly.
In this section, we study more
details of possible experimental probes.
The photon signal at a hadron collider is very clear, so we simply use
photon information to obtain sensitivities for the
$gg \rightarrow \gamma \gamma$ channel.
The event spectrum for the $gg \rightarrow \gamma \gamma$ process
has been studied carefully in \cite{Ellis:2018cos}, and we discuss the updated sensitivity of 
$gg \rightarrow \gamma \gamma$ in \gsec{sec:gg2AA}.

\begin{figure}[t]
\centering
\includegraphics[width=0.48\textwidth]{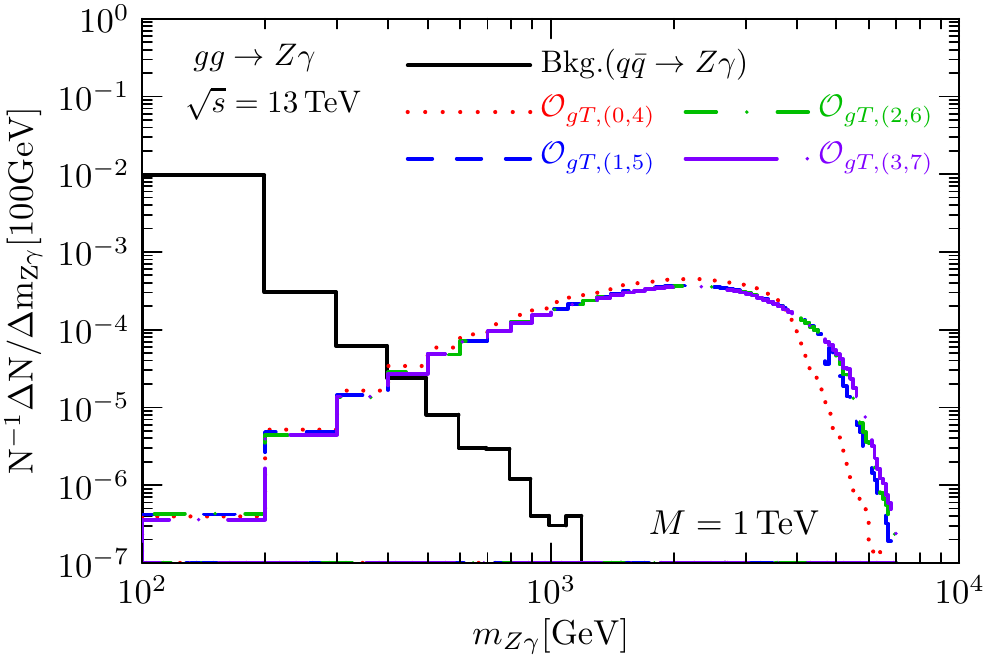}
\quad
\includegraphics[width=0.48\textwidth]{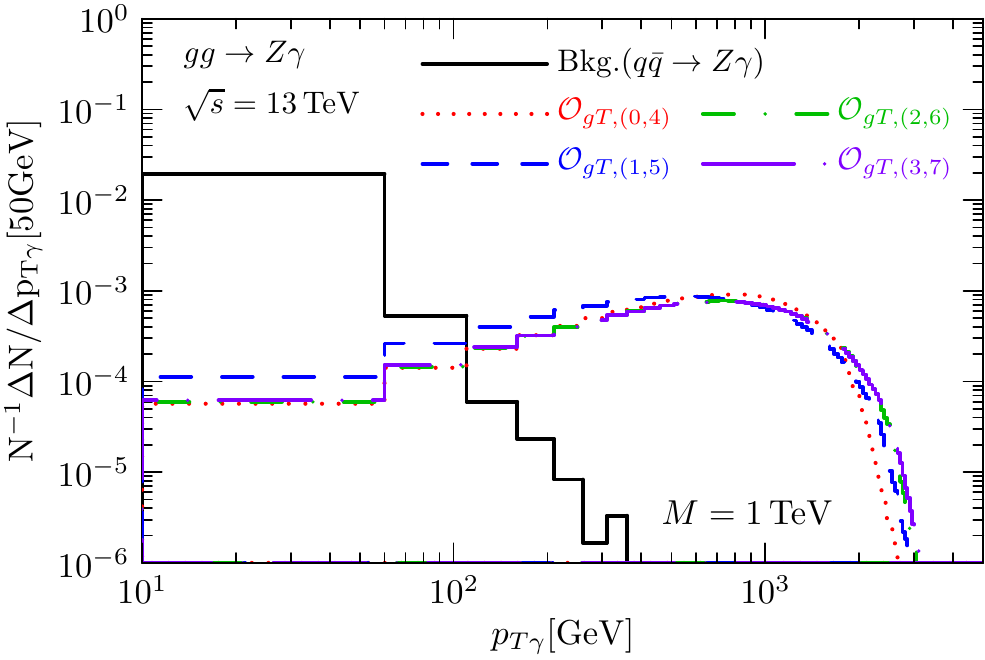}
\caption{\it The normalized event spectra due to the background (black
solid line) and gQGC operators (colorful non-solid lines) as functions
of the $Z\gamma$ invariant mass $m_{Z\gamma}$ (left)
and the photon transverse momentum $p_{T \gamma}$ (right)
at the LHC with $\sqrt{s} = 13 \, \tev$. As seen more clearly in the left panel,
the total cross sections have been regulated to respect the unitarity constraint.}
\label{fig:spectrum}
\end{figure}

The normalized event spectra for $gg \to Z \gamma$ due to the
various dimension-8 operators at
the LHC with $\sqrt s = 13$\,TeV are shown in
\gfig{fig:spectrum}. The major differences between \gfig{fig:xsec}
and \gfig{fig:spectrum} are due to the gluon parton distribution
function (PDF) in proton, which suppresses the high-energy tails of the event
spectra. This makes the difference between the gQGC signals and the SM background even
more prominent, and reduces the sensitivity to the treatment of the unitarity constraint. 
Spectra in the $Z \gamma$ invariant mass
$m_{Z \gamma}$ are shown in the left panel of \gfig{fig:spectrum} and the photon transverse momentum
$p_{T \gamma}$ distributions are shown in the right panel.~\footnote{Analogous $m_{\gamma \gamma}$ distributions
for $gg \to \gamma \gamma$ scattering were shown in Fig.~2 of~\cite{Ellis:2018cos}.}
Comparing the two panels of \gfig{fig:spectrum}, we see that the $p_{T \gamma}$
spectra grow more slowly. In the cases of charged-fermion final states, $Z \rightarrow \ell^+ \ell^-$
and $q \bar q$, the invariant mass spectrum contains more information,
by virtue of the $p^8 \sim s^4$ energy dependence. However, in the case of the decays
into neutrinos, $Z \rightarrow \nu \bar \nu$, the
only reliable experimental information is provided by the photon transverse momentum $p_{T \gamma}$.

We study individually the $Z$ boson
decays into various fermion-antifermion pairs: 
decays into charged leptons, $Z \rightarrow \ell^+ \ell^-$, in \gsec{sec:Zlep};
decays into neutrinos, $Z \rightarrow \bar \nu \nu$, in \gsec{sec:Znu};
and decays into quarks, $Z \rightarrow q \bar q$, in \gsec{sec:Zq}.
The prospects for all these channels at
the future hadron colliders (HE-LHC, FCC-hh, and SppC) are studied in
\gsec{sec:future}. 
The event rates in this paper are simulated with {\tt MadGraph5}
\cite{Alwall:2014hca} with model files prepared by {\tt Feynrules}
\cite{Alloul:2013bka} in the UFO format \cite{Degrande:2011ua}
\gred{for the leading order (LO) results.
NLO corrections to the background at $\sqrt{s}=13~{\rm TeV}$
are taken into account by an overall $K$-factor that is extracted
by comparing our LO calculations with the NLO calculations given 
in the experimental papers (the $pp\to\gamma\gamma$ process reported
in \cite{ATLAS:2021mbt}, the $pp\to Z\gamma$ processes in \cite{ATLAS:2019gey}
for the leptonic channel and \cite{ATLAS:2018nci} for the invisible
decay channel as well as \cite{ATLAS:2018sxj} for the hadronic channel). 
For future colliders, we have used the same overall $K$-factor.}

We adopt the same binning for the prospective signals, $N^{\rm sig}_i$, as has been used for the
corresponding experimental data points, $N^{\rm data}_i$.
On the basis of these numbers, we use the following $\chi^2$ function to evaluate
the sensitivity to the dimension-8 gQGC operators:
\begin{equation}
  \chi^2
=
  \sum_i
\left[
  \frac {N^{\rm data}_i - F N^{\rm bkg}_i \Pi_k (1 + \bar \sigma_{k,i} + f_k \sigma_{k,i}) - N^{\rm sig}_i}
        {\sqrt{N^{\rm data}_i}}
\right]^2
+
  \sum_k f^2_k.
\label{eq:chi2}
\end{equation}
In addition to the cut-off scale $M_i$ that enters implicitly through
the signal event rates $N^{\rm sig}(M_i)$, the $\chi^2$ function also
contains an overall normalization $F$ and multiple nuisance parameters
$f_k$ to account for the various systematics published in the
experimental papers. For each systematical error, both the central
value $\bar \sigma_i$ and the deviations $f_k \sigma_{k,i}$ have been
taken into account. Assuming quadratic dependences on the nuisance parameters
$F$ and $f_k$, the $\chi^2$ minimization can be done
analytically \cite{Ge:2012wj,Ge:2016zro}. After marginalizing over
$F$ and $f_k$, the resulting $\chi^2(M_i)$ as function of a single
parameter provides the sensitivity to the cut-off scale $M_i$.
The same procedure has been followed for the SM background, so as to calculate
$\chi^2_{\rm min}$. The square root of the difference,
$\Delta \chi^2(M_i) \equiv \chi^2(M_i) - \chi^2_{\rm min}$, directly
gives the significance that we show below. We generally quote results at the
95\% C.L., corresponding to $\Delta \chi^2 = 3.84$.
\gred{The sensitivities shown in this
paper are obtained for each of the 8 gQGC operators separately and for the
BISM combination. The constraints on the individual operator
coefficients would in general be weaker if all the operators were 
included simultaneously and one marginalized over the other operator
coefficients.}

\gred{Below we apply this $\chi^2$ analysis to the ATLAS
data for various channels.} To guarantee that the
quadratic $\chi^2$ function is a good enough approximation, only those bins with
at least 5 events are included in \geqn{eq:chi2}. This provides a very
conservative sensitivity estimation, since beyond the end points of the
SM background (or the experimental data), the signal is still
growing and essentially free of background. If a more sophisticated
data analysis were employed, the sensitivity could be further improved.
We leave this to our experimental colleagues.

\subsection{$pp\to\gamma\gamma$ at the LHC}
\label{sec:gg2AA}

\begin{figure}[t]
\centering
\includegraphics[width=0.48\textwidth]{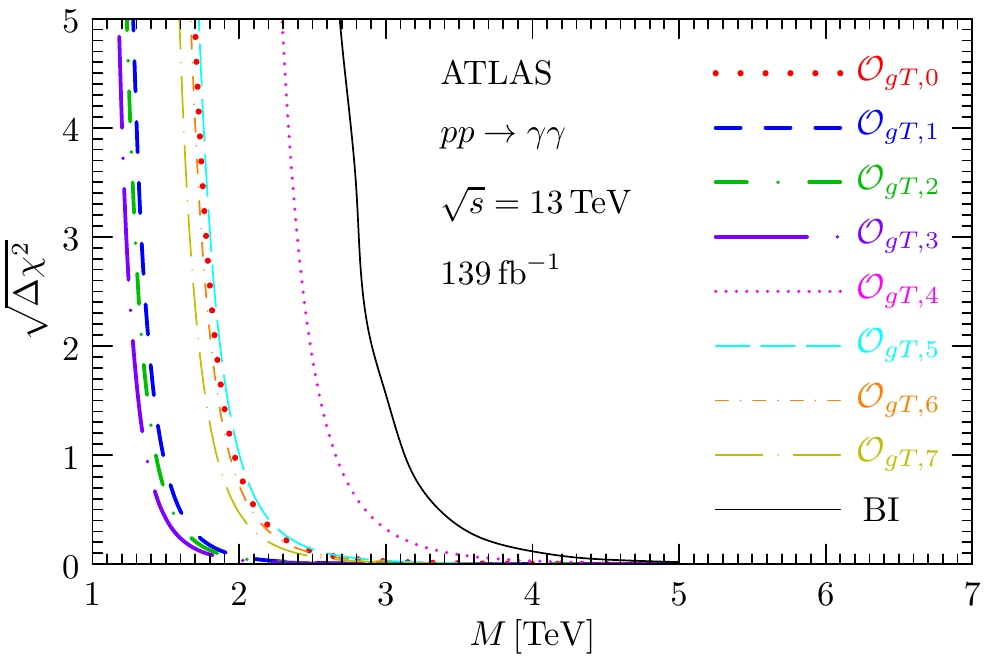}
\quad
\includegraphics[width=0.48\textwidth]{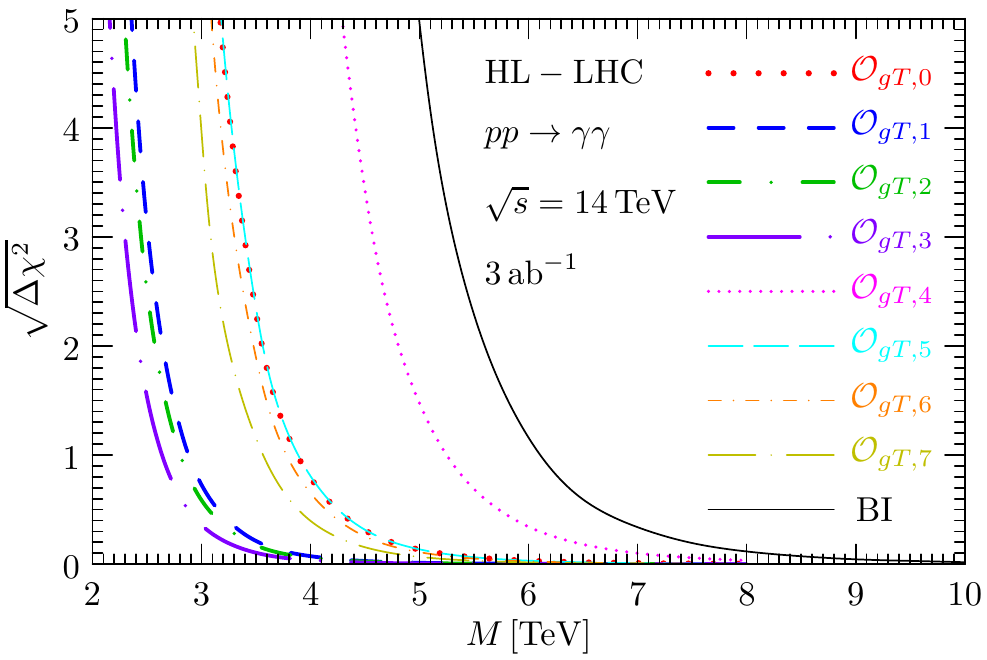}
\caption{\it The significances $\sqrt{\Delta \chi^2}$ for probes of the
\gred{individual}
dimension-8 gQGC operators ${\cal O}_{gT,i}$ and their Born-Infeld
combination in Eq.\,(\ref{LBISM}) at the LHC as functions of the cut-off scales $M_i$.
These results are obtained from the $gg \rightarrow \gamma \gamma$ reaction at 
$\sqrt{s} = 13 \, \tev$ with integrated luminosity $\call=139\fb^{-1}$
observed by ATLAS (left) and $\sqrt{s} = 14 \, \tev$ with $\call=3\ab^{-1}$
at HL-LHC (right).}
\label{fig:chi2AA}
\end{figure}

We first update the analysis of \cite{Ellis:2018cos} using the ATLAS measurement of
the diphoton \gred{invariant mass $m_{\gamma \gamma}$ spectrum} using 36.7\,fb$^{-1}$ to include 139\,fb$^{-1}$
\cite{ATLAS:2021mbt}, which provides improved sensitivities to dimension-8 operators.
The left panel of \gfig{fig:chi2AA} displays the updated results from
analyzing the ATLAS data on isolated $\gamma \gamma$ production. 
Because of the different electroweak mixing angle factors in \geqn{WWBB},
the $\sum_i W^i W^i$ operators $\mathcal O_{gT, (0,1,2,3)}$ with
coefficients $s^2_w$ are constrained more weakly than their $BB$
counterparts $\mathcal O_{gT, (4,5,6,7)}$, which have factors $c^2_w$ instead.
The 95\% C.L. lower limits on the mass scales $M_i$ reach
$\simeq 1.3$\,TeV for $\mathcal O_{gT,(1,2,3)}$,  $\simeq 1.8$\,TeV
for $\mathcal O_{gT,(0,5,6,7)}$, and $\simeq 2.5$\,TeV for
$\mathcal O_{gT,4}$.
Since the improvement comes principally from the updated luminosity, further
enhancement in the collision energy and luminosity at the LHC can
push the limits to higher values.
The right panel of \gfig{fig:chi2AA} shows the prospects for LHC running
at the design energy $\sqrt{s} = 14 \, \tev$ with integrated
luminosity $\call=3\ab^{-1}$ \cite{CidVidal:2018eel}.
The constraints at the same 95\% C.L. can improve by amounts 
$\sim 1$ to $2\, \tev$.

The solid black line in \gfig{fig:chi2AA} represents the sensitivity
to the Born-Infeld ${\cal O}_{gT, BI}$ combination of the
gQGC operators in \geqn{LBISM}. This linear combination of 6 individual
gQGC operators yields an event rate that is much larger than any individual operator, as
seen in \gfig{fig:xsec}. Consequently, the 95\% CL lower limit on the BISM mass scale
reaches $M_{BI} \gtrsim 5$\,TeV. This result is strong enough
to impose a significant constraint on the separation between branes
in some M-theory inspired models that address the electroweak hierarchy problem.

We emphasize that we are working to quadratic order in the
dimension-8 operator coefficients, unlike many analyses of dimension-6 operators at
the LHC. Also, our analysis is reliant on values of $\sqrt{\hat s}$ that are below those
where the unitarity constraints become important. However, we also note that
in some instances the values of $\sqrt{\hat s}$ contributing to our analysis
are comparable to (or even exceed) the magnitudes of the constraints on new-physics
scale $M_i$ that we derive. In such cases, the validity of our constraints depends
on the magnitude of the new-physics coupling. They would be most reliable if the dimension-8 operator were generated 
by some strongly-coupled new physics at a scale $\Lambda$ with coupling strength $4 \pi$, 
in which case $\Lambda = \sqrt{4 \pi} M$ would generally exceed the values of $\sqrt{\hat s}$
that provide the constraints in our analysis.

The CMS and TOTEM collaborations have recently searched for
exclusive diphoton production in proton-proton collisions and set constraints on the dimension-8 quartic
photon couplings $\zeta_1 F_{\mu \nu} F^{\mu \nu} F_{\rho \sigma} F^{\rho \sigma}$
and $\zeta_2 F_{\mu \nu} F^{\mu \rho} F_{\rho \sigma} F^{\sigma \nu}$
\cite{TOTEM:2021kin}.
Their result is
a two-dimensional contour in the plane of the coefficients
$\zeta_1$ and $\zeta_2$, in the absence of theoretical constraints.
As demonstrated earlier, the Born-Infeld QED extension
in \geqn{BornInfeld} naturally correlates the two purely photon QGC
operators with $\zeta_1 = \zeta_2 = 1 / {32 M^4}$.
Thus the 95\% sensitivity of the CMS-TOTEM data becomes
$M_{\rm BI} \geq 670\,$GeV. The CMS and TOTEM 
sensitivity is limited by the fact that it uses $\gamma \gamma \rightarrow \gamma \gamma$ scattering
with photons in both the initial and final states, leading to a 
weaker constraint than the $gg \rightarrow \gamma \gamma$
process at the same collider for either individual operator or the
combined BISM operator. This confirms the advantage of considering the
$gg \rightarrow \gamma \gamma$ channel at hadron
colliders.

\subsection{$pp \to Z(\ell^+ \ell^-)\gamma$ at the LHC} 
\label{sec:Zlep}

The ATLAS Collaboration has searched for new physics in the 
$Z(\ell^{+}\ell^{-})\gamma$ final state
with an integrated luminosity of $139\fb^{-1}$ \cite{ATLAS:2019gey}
at $\sqrt{s}=13~\tev$.
Here we re-interpret the results with electron ($Z \rightarrow e^+ e^-$) and muon
($Z \rightarrow \mu^+ \mu^-$) final states as constraints on the gQGC operators. Both the photon and
the charged leptons are observable. 
The event selection required the 
photon to have pseudo-rapidity in the range
$|\eta^{\gamma}| < 2.37$ and transverse energies $E_T^\gamma > 30\,\gev$.
We use the ATLAS data with tight photon identification, which had an identification
efficiency ranging from $82-85\%$ for $E_{T}^{\gamma} \approx 30\,\gev$ 
to $90-98\% $ for $E_{T}^{\gamma} > 100\,\gev$. The
electrons and muons are required to have pseudo-rapidities
$|\eta^{\ell}| < 2.47$ and transverse momenta $p_T^\ell > 25\,\gev$.
The identification efficiency for charged leptons is about
$80\%$ ($93\%$) for $p_T \approx 25\,\gev$ ($100\,\gev$). 
Because of geometrical limitations, the transition region between the
barrel and endcap $(1.37 < |\eta| < 1.52)$ is excluded for both the photon
and the charged leptons.

The dominant instrumental backgrounds for the $Z(\ell^{+}\ell^{-})\gamma$
final states are $Z+{\rm jets}$, pile-up events with one photon,
$t\bar{t}\gamma$ and $\tau^{+}\tau^{-}\gamma$, 
as well as double vector boson production with or without an
isolated photon \cite{ATLAS:2019gey}. It was estimated that the
combination of these backgrounds contributes about
$18\%$ of the signal process $q\bar{q} \to Z(\ell^{+}\ell^{-})\gamma$
in the fiducial phase space. For simplicity, we take the {\tt Sherpa}
simulation of the background from the ATLAS paper \cite{ATLAS:2019gey}
and assign an overall normalization factor $F$ when fitting the
experimental data as explained in the discussion of \geqn{eq:chi2}.

\begin{figure}[t]
\centering
\includegraphics[width=0.48\textwidth]{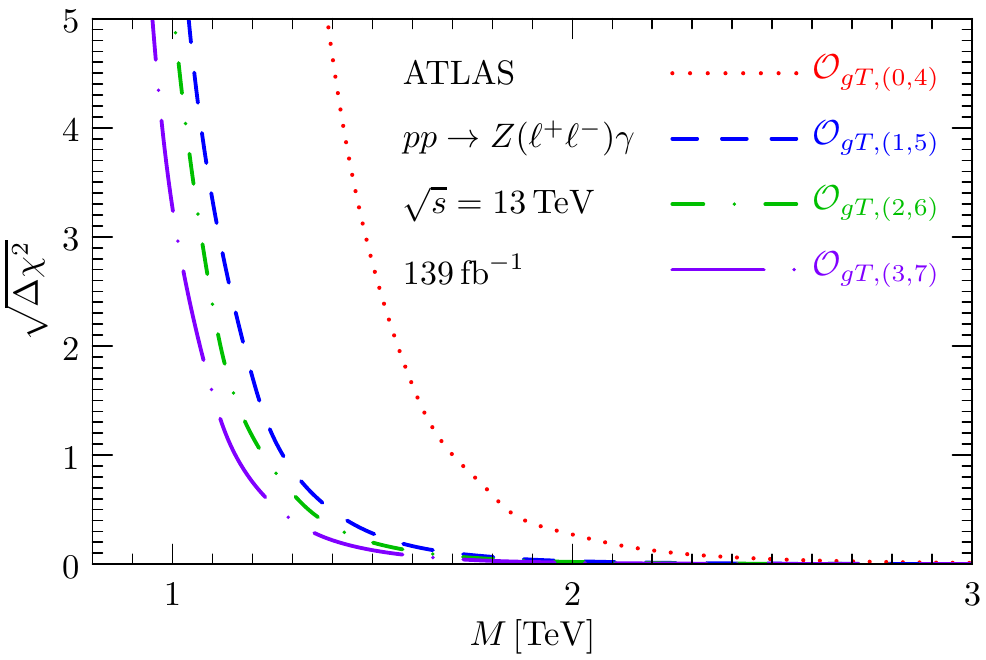}
\quad
\includegraphics[width=0.48\textwidth]{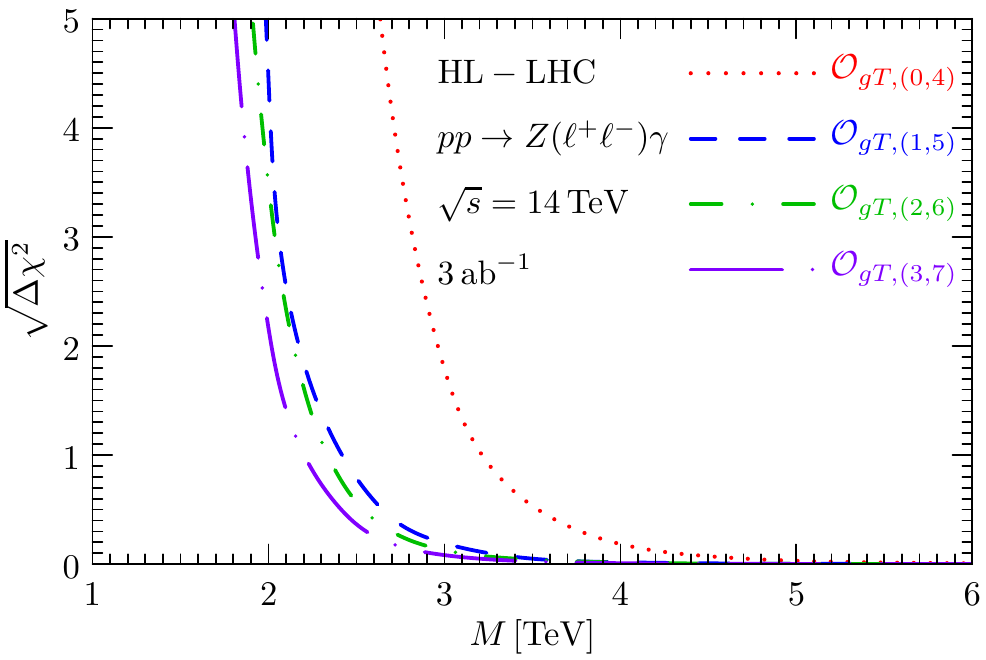}
\caption{\it The significances $\sqrt{\Delta \chi^2}$ for probes of the 
\gred{individual} dimension-8
gQGC operators ${\cal O}_{gT,(i, i+4)}$, where $i = 0,1,2,3$, from the
$Z(\ell^+ \ell^-) \gamma$ channel as functions of the cut-off scales $M_i$.
The left panel shows the sensitivities obtained using the current
LHC data with $\sqrt{s} = 13\,\tev$ and luminosity $\call=139\fb^{-1}$,
and the right panel shows the estimated sensitivities for the future HL-LHC with
$\sqrt{s} = 14\,\tev$ and $\call=3\ab^{-1}$.}
\label{fig:ZA:2L:LHC13}
\end{figure}

The left panel of \gfig{fig:ZA:2L:LHC13} shows the sensitivities
\gred{obtained from fitting the observed $Z(\to \ell^+\ell^-) \gamma$
invariant mass $m_{Z \gamma}$ spectrum} at ATLAS \cite{ATLAS:2019gey}
for probing the gQGC operators. Since the contributions of
$G G \sum_i W^i W^i$ and $G G B B$ operators with the same Lorentz
structures have identical EW mixing factors for the $Z \gamma$
combination, except for a relative sign, see \geqn{WWBB},
the sensitivities to $\mathcal O_{gT,i}$ and
$\mathcal O_{gT,i+4}$ (i = 0, 1, 2, 3) are pairwise identical.
Hence we plot only 4 curves, for $\mathcal O_{gT,(0,4)}$
(red), $\mathcal O_{gT,(1,5)}$ (blue), $\mathcal O_{gT,(2,6)}$ (black),
and $\mathcal O_{gT,(3,7)}$ (green), respectively.
The 95\% C.L. lower limits on the new physics scales $M_i$ can
reach $\simeq 1$\,TeV for $\mathcal O_{gT,(1,2,3,5,6,7)}$ and
$\simeq 1.7$\,TeV for $\mathcal O_{gT,(0,4)}$. 
For comparison, the right panel of \gfig{fig:ZA:2L:LHC13} shows the prospects at the HL-LHC
with $\sqrt{s} = 14\,\tev$ and integrated luminosity $\call=3\ab^{-1}$.
We see that the lower limits are enhanced by roughly $1\,\tev$.~\footnote{We recall that
the $Z \gamma$ final state does not provide any constraint
on the Born-Infeld scale, since the contributions of the $W^3 W^3$ and $BB$
operators cancel with each other as explained below \geqn{WWBB}.}

With three detectable particles in the final state, much more
information can in principle be extracted beyond just the invariant mass, pseudo-rapidity,
and transverse momentum. For example, ATLAS also analyzed the
angular distribution of the
$\ell^+ \ell^-$ pair relative to the scattering plane.
We discuss in \gsec{sec:improvements} potential improvements of the cross-section analysis 
using spin and angular correlation measurements.

\subsection{$pp \to Z(\bar \nu \nu)\gamma$ at the LHC}
\label{sec:Znu}

\begin{figure}[t]
\centering
\includegraphics[width=0.48\textwidth]{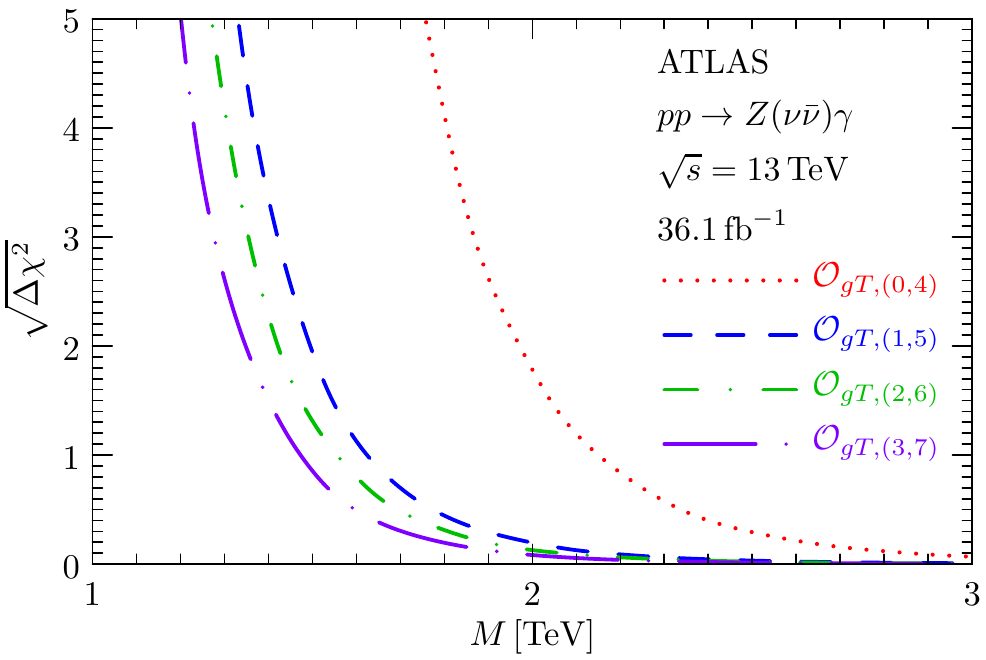}
\quad
\includegraphics[width=0.48\textwidth]{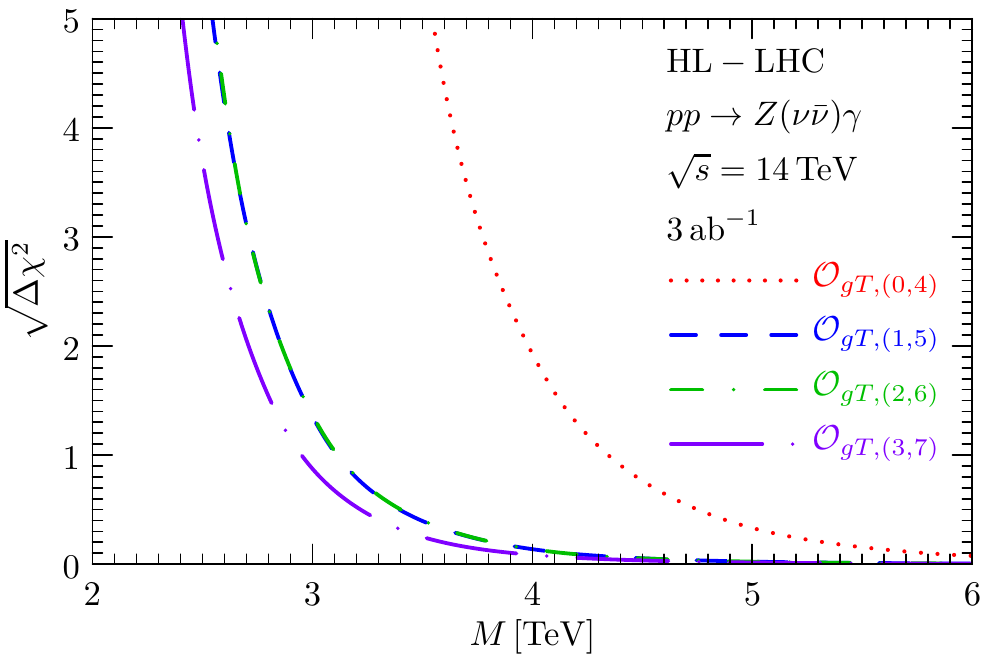}
\caption{\it The significances $\sqrt{\Delta \chi^2}$ for probes of the 
\gred{individual} dimension-8
gQGC operators ${\cal O}_{gT,(i, i+4)}$, where $i = 0,1,2,3$, from the
$Z(\nu\bar\nu) \gamma$ channel as functions of the cut-off scales $M_i$.
The left panel shows the sensitivities obtained using the current
LHC data with $\sqrt{s} = 13\,\tev$ and luminosity $\call=36.1\fb^{-1}$
while the right panel shows the estimated sensitivities for the future HL-LHC with
$\sqrt{s} = 14\,\tev$ and $\call=3\ab^{-1}$.} 
\label{fig:ZA:2N:LHC13}
\end{figure}


The detection of the $Z(\bar \nu \nu)\gamma$ final state differs from those of the
diphoton channel and the $gg \rightarrow Z (\ell^+ \ell^-) \gamma$ mode,
since only the photon is observable. This renders impossible the reconstruction of the invariant mass
spectrum that is the optimal choice for seeing the momentum dependence
of the dimension-8 gQGC operators. The only usable information
is the photon energy/momentum vector. Although this can readily be measured, the information cannot be fully
used since the initial momenta of the colliding gluons in the beam protons are unknown. Because of this uncertainty,
one can only utilize the transverse momentum spectrum depicted in the
right panel of \gfig{fig:spectrum}. Although not optimal, it still
possesses the feature that the event rate keeps growing with momentum
until the unitarity saturation is encountered and the gluon PDF
suppression finally dominates.

The ATLAS measurement of $Z (\to \bar \nu \nu) + \gamma$ uses an
integrated luminosity of ${\cal L} = 36.1$\,fb$^{-1}$ at 13\,TeV
\cite{ATLAS:2018nci}. The event selection uses the following experimental cut
on the transverse energy: $E_T^\gamma$, $E^{\rm miss}_T > 150$\,GeV. 
Similarly to the charged lepton case, the photon
pseudo-rapidity is in the range $|\eta^\gamma| < 2.37$, but
excluding $1.37 < |\eta^\gamma| < 1.52$ to avoid the gap between
the detector barrel and end-cap regions. The transverse missing momentum
$p^{\rm miss}_T$ needs not be opposite to the photon transverse
momentum $p^\gamma_T$, due to parton shower from the initial gluons.
Requiring the azimuthal angular difference
$\Delta \phi(p_\gamma, p_T^{\rm miss}) > \pi/2$
further suppresses the background.

The dominant instrumental background for the
$Z(\bar \nu \nu)\gamma$ final state comes from $W \gamma$
associated production with the $W$ decaying leptonically but the charged
lepton from $W \rightarrow \ell \nu$ not being identified. In addition,
jets $+ \gamma$ and other events with either an electron or a jet
being misidentified as a $\gamma$ are possible sources of background
\cite{ATLAS:2018nci}. These backgrounds can amount to about
$68 \, (58)\%$ of the inclusive (exclusive) SM background
$q\bar{q} \to Z(\bar \nu \nu)\gamma$ in the fiducial phase space,
i.e., the cross section with $\ge 0 \, (0)$ additional jets.

The sensitivities from analytical $\chi^2$ fits to the different gQGCs obtained using \geqn{eq:chi2}
are shown in the left panel of \gfig{fig:ZA:2N:LHC13}. Although the
$pp \rightarrow Z (\bar \nu \nu) \gamma$ channel is not the optimal
one, the lower bound on the cut-off scale $M$ at 95\% C.L. can still
reach around 1.4\,TeV for $\mathcal O_{gT, (1,2,3,5,6,7)}$ and 1.9\,TeV
for $\mathcal O_{gT,(0,4)}$. One reason is that the
$Z \rightarrow \bar \nu \nu$ mode includes all three
neutrinos whereas only electron and muon modes are used for the charged
lepton mode. Moreover, the $Z$ decay branching ratio into a single
neutrino is roughly double that of its charged lepton counterpart, so the
combined $Z \to \bar \nu \nu$ branching ratio is about three times larger than
the combined $Z \to \ell^+ \ell^-$ branching ratio.
The right panel of \gfig{fig:ZA:2N:LHC13} shows the prospects at the
HL-LHC with its higher energy and luminosity. The lower bounds are
improved by $\gtrsim 1$~TeV.

\subsection{$pp \to Z(q\bar{q})\gamma$ at the LHC}
\label{sec:Zq}

\begin{figure}[t]
\centering
\includegraphics[width=0.325\textwidth]{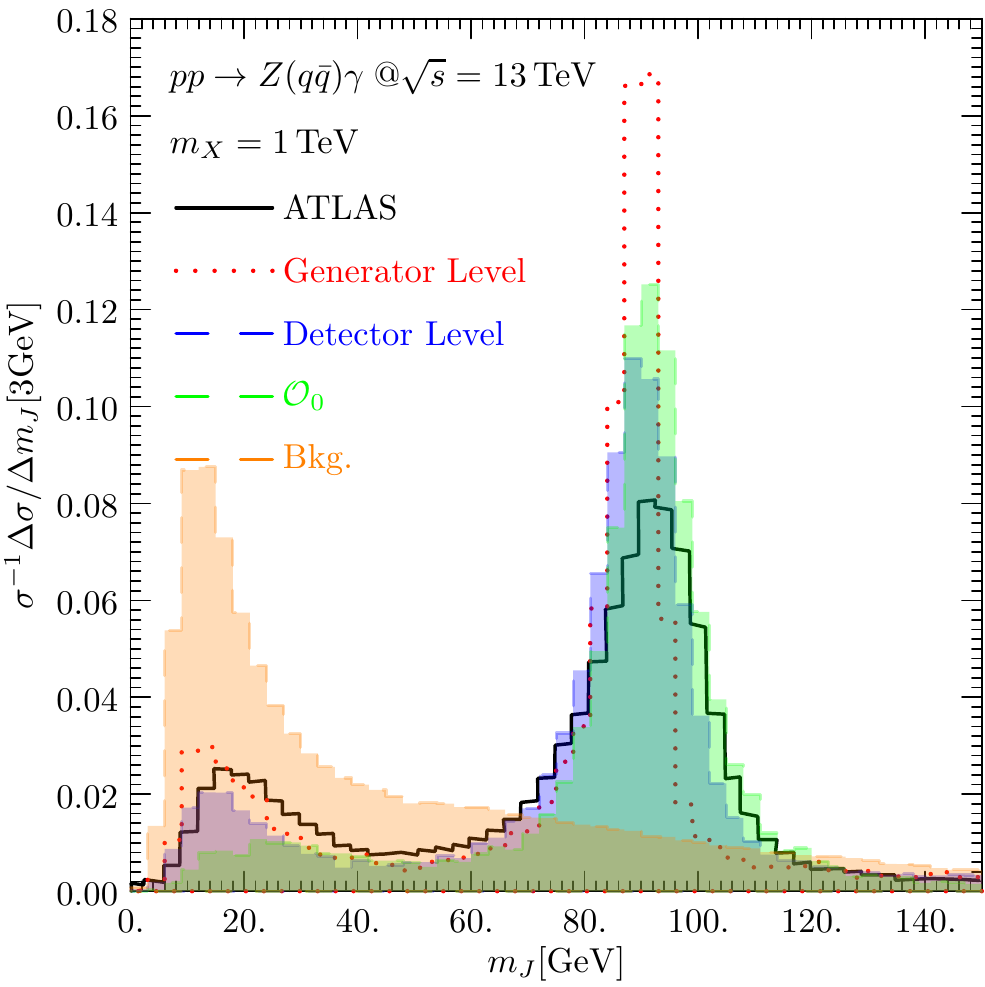}
\hfill
\includegraphics[width=0.32\textwidth]{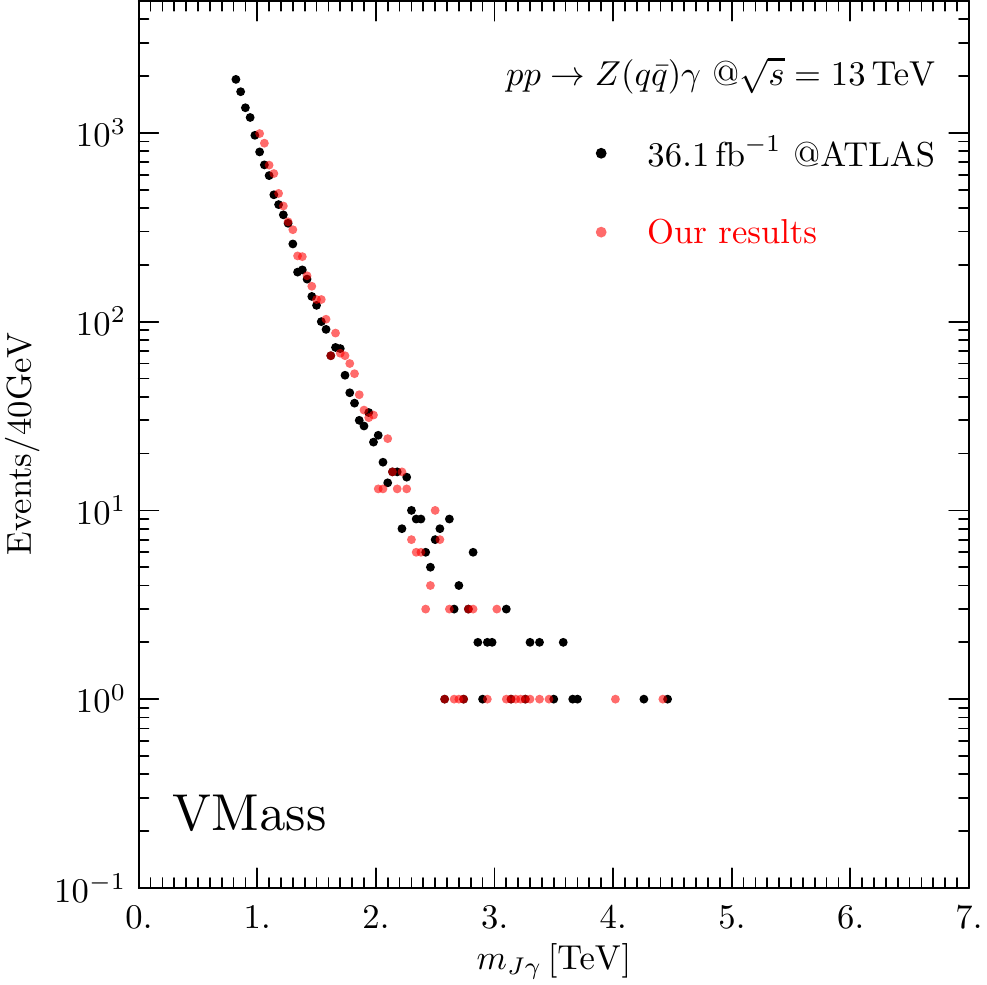}
\hfill
\includegraphics[width=0.32\textwidth]{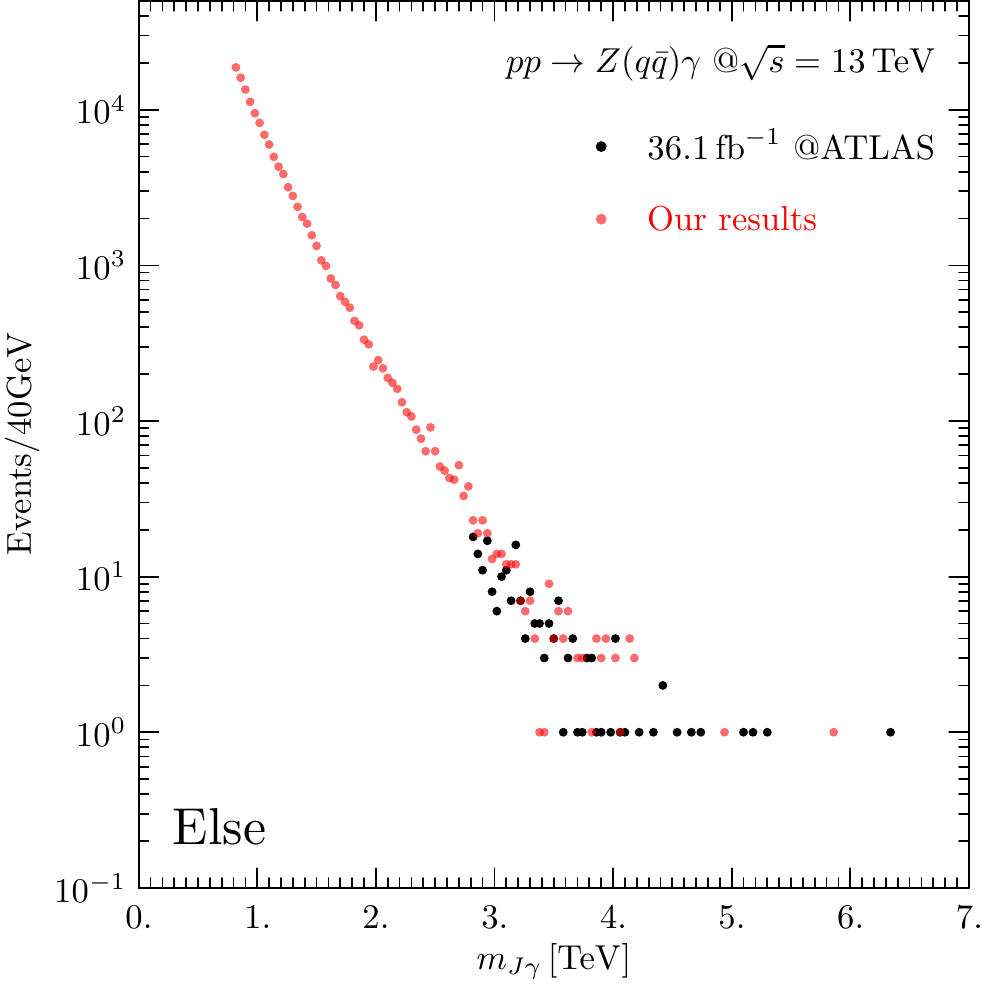}
\caption{\it Validations of our simulation of the hadronic decay mode
$Z\to q\bar{q}$, assuming a scalar resonance $X$ with mass $m_{X}=1\,\tev$.
The left panel shows the invariant mass distribution
of a fat jet emerging from the decay of a highly-boosted $Z$-boson while
the middle and right panels compare distributions of the invariant
mass $m_{J\gamma}$ for the VMass and Else categories.} 
\label{fig:ZA:2J:LHC13:Valid}
\end{figure}

The leptonic modes discussed in the previous sections have
very clean signals at the hadron colliders. However, although the hadronic
decay modes of the $Z$ boson have more complicated
backgrounds, the larger branching ratio provides some advantages.
The ATLAS Collaboration has measured the $Z (\to \bar q q) \gamma$
\gred{$m_{Z \gamma}$ invariant mass spectrum}
with an integrated luminosity of ${\cal L} = 36.1$\,fb$^{-1}$
at 13\,TeV \cite{ATLAS:2018sxj}.
The experimental analysis first selected events containing a
hadronic jet with $p_T > 200$\,GeV and $|\eta| < 2.0$ in addition to
a photon with $p_T^\gamma > 250$\,GeV and $|\eta| < 1.37$.
The events were then divided into four  categories \cite{ATLAS:2018sxj}:
1) The BTAG category of events in which the two leading
track-jets associated with a large-radius jet candidate satisfy
the jet mass requirement for $b$-tagging, with a 
$Z \to \bar q q$ identification efficiency of 3 to 4\%;
2) The D2 category that is composed of events satisfying combined
jet mass and substructure discriminant requirements with an identification efficiency of 20 to 28\%;
3) The VMass category containing events that pass the jet mass selection
but fail to enter either of the two previous categories (identification efficiency 24 to 36\%); and
4) The Else category containing the remaining events that pass the baseline selection 
(identification efficiency 40 to 50\%). Roughly speaking,
the event rates in the above four categories increase by an order of magnitude
going from category 1) to category 4).
\gfig{fig:ZA:2J:LHC13:Valid} shows some validation results from our simulation obtained using
{\tt Pythia8} \cite{Sjostrand:2014zea} for parton shower effects and 
{\tt Delphes3} \cite{deFavereau:2013fsa} for detector effects,
as well as {\tt FastJet}\cite{Cacciari:2011ma}, {\tt ExRootAnalysis}\cite{Alwall:2014hca} and
{\tt Root} \cite{Brun:1997pa} for data analysis. The invariant mass
distribution in the left panel reproduces adequately the $Z$ boson
peak and a lower background peak around 20\,GeV. Since the event rate
is dominated by the VMass and Else categories, we only show these
two in the middle and right panels. Our simulations are reasonably
consistent with the ATLAS data points.

\begin{figure}[t]
\centering
\includegraphics[width=0.48\textwidth]{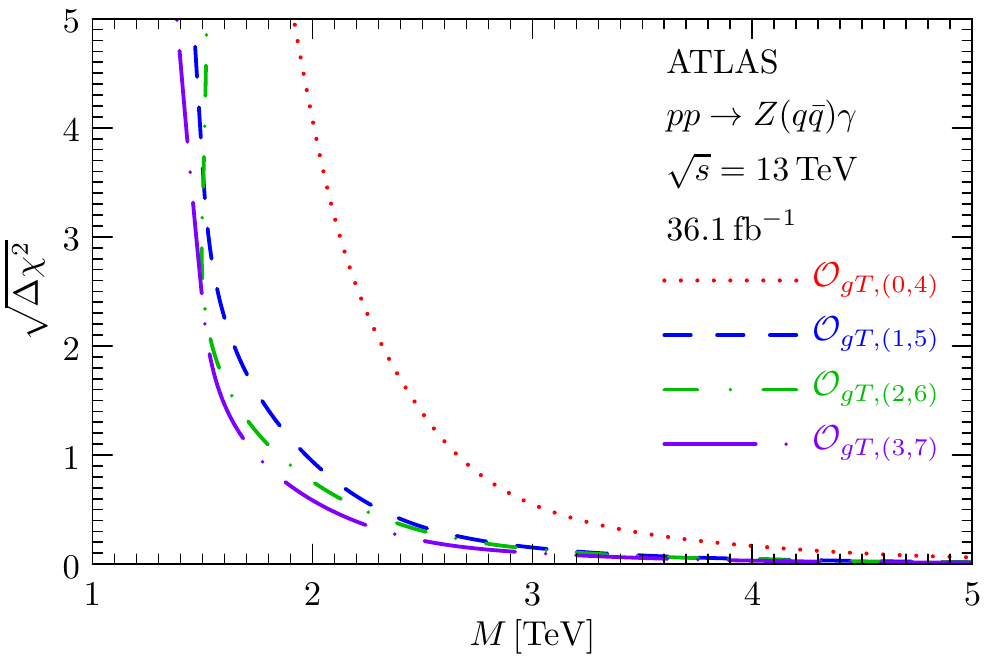}
\quad
\includegraphics[width=0.48\textwidth]{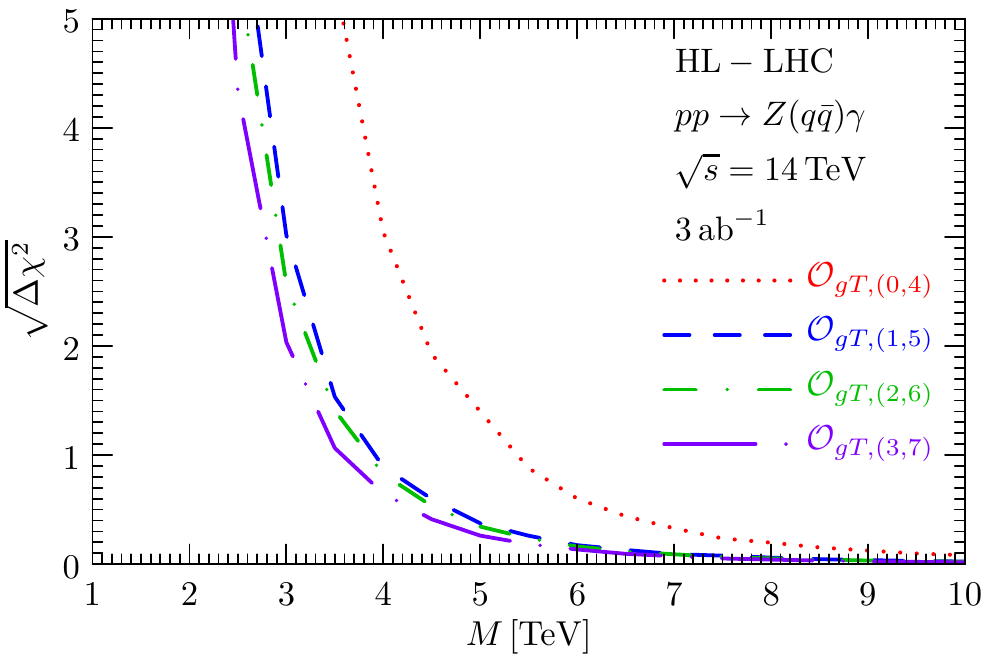}
\caption{\it The significances $\sqrt{\Delta \chi^2}$ for probes of the 
\gred{individual} dimension-8
gQGC operators ${\cal O}_{gT,(i, i+4)}$, where $i = 0,1,2,3$, from the
$Z(q\bar q) \gamma$ channel as functions of the cut-off scales $M_i$.
The left panel shows the sensitivities obtained using the current
LHC data with $\sqrt{s} = 13\,\tev$ and luminosity $\call=36.1\fb^{-1}$ while
the right panel shows the estimated sensitivities for the future HL-LHC with
$\sqrt{s} = 14\,\tev$ and $\call=3\ab^{-1}$.} 
\label{fig:ZA:2J:LHC13}
\end{figure}

The sensitivities for the $pp \rightarrow Z(q \bar q) \gamma$
channel at the LHC are shown in \gfig{fig:ZA:2J:LHC13} for both
the current LHC data (left panel) and the future HL-LHC (right panel).
Although the hadronic modes are much more difficult to detect,
the 95\% C.L. lower bound still reaches around 1.5\,TeV for
$\mathcal O_{gT, (1,2,3,5,6,7)}$ and 2.3\,TeV for
$\mathcal O_{gT, (0,4)}$, which are even stronger than the neutrino
mode. Further improvement by at least another 1\,TeV should be possible
at HL-LHC.
In addition to the pseudo-rapidity and transverse momentum, the
angular distribution of the $\bar q q$ pair relative to the
scattering plane can also be measured. However, this was not
done in the ATLAS analysis \cite{ATLAS:2018sxj}. A discussion
of possible future improvements using this information is given in \gsec{sec:improvements}.

\section{Prospective Sensitivities at Future Hadron Colliders}
\label{sec:future}

In addition to the existing LHC and its upgrade to HL-LHC,
other hadronic colliders are being proposed, including
HE-LHC \cite{CidVidal:2018eel}, FCC-hh \cite{FCC:2018vvp}, 
and SppC \cite{CEPC-SPPCStudyGroup:2015csa},
whose collision energies range from 25\,TeV to 50\,TeV and 100\,TeV.
As discussed in \gsec{sec:xsec}, because of its strong momentum
dependence, the gQGC signal increases very rapidly with the
colliding energy. This gives future higher-energy hadron
colliders great advantages for probing the gQGC operators.
In addition, future hadron colliders typically have much
larger luminosity than the current LHC. We assume a universal figure of
$\mathcal L = 20\,\ab^{-1}$ for the various higher-energy colliders
and higher $p_T$ cutoffs than for the LHC, as shown in 
\gtab{tab:hptc}, but retain the same scattering-angle cuts as for the LHC.
\begin{table}[h]
\renewcommand\arraystretch{1.44}
\begin{center}
\begin{tabular}{c c c c}
{\rm Process}   & ~~~25~TeV~~~ & ~~~50~TeV~~~ & ~~~100~TeV~~~
%
%
\\\hline\hline
$pp\to \gamma\gamma$ 
& 
$p_{T,\gamma} = 100 \, \gev$ 
& 
$p_{T,\gamma} = 150 \, \gev$ 
& 
$p_{T,\gamma} = 300 \, \gev$ 
\\[2mm]\hline
$pp\to Z(\ell^{+}\ell^{-})\gamma$ 
& 
\begin{tabular}{c} $p_{T,\gamma} = \, 60 \, \gev$ \\ $p_{T,\ell} = 50 \, \gev$ \end{tabular}
& 
\begin{tabular}{c} $p_{T,\gamma} = 100 \, \gev$ \\ $p_{T,\ell} = 80 \, \gev$ \end{tabular}
& 
\begin{tabular}{c} $p_{T,\gamma} = 200 \, \gev$ \\ $p_{T,\ell} = 160 \, \gev$ \end{tabular}
\\[2mm]\hline
$pp\to Z(\nu\bar\nu)\gamma$ 
& 
\begin{tabular}{c} $p_{T,\gamma} = 250 \, \gev$ \\ ${p\!\!\!\slash}_{T} = 250 \, \gev$ \end{tabular}
& 
\begin{tabular}{c} $p_{T,\gamma} = 450 \, \gev$ \\ ${p\!\!\!\slash}_{T} = 500 \, \gev$ \end{tabular}
& 
\begin{tabular}{c} $p_{T,\gamma} = 1000 \, \gev$ \\ ${p\!\!\!\slash}_{T} = 1000 \, \gev$ \end{tabular}
\\[2mm]\hline
$pp\to Z(q\bar{q})\gamma$ 
& 
$p_{T,\gamma} = 400 \, \gev$
& 
$p_{T,\gamma} = 800 \, \gev$ 
& 
$p_{T,\gamma} = 1600 \, \gev$
%
%
\end{tabular}
\caption{\it Transverse momentum cuts at higher-energy colliders}
\label{tab:hptc}
\end{center}
\end{table}

\begin{figure}[t]
\centering
\includegraphics[width=0.98\textwidth]{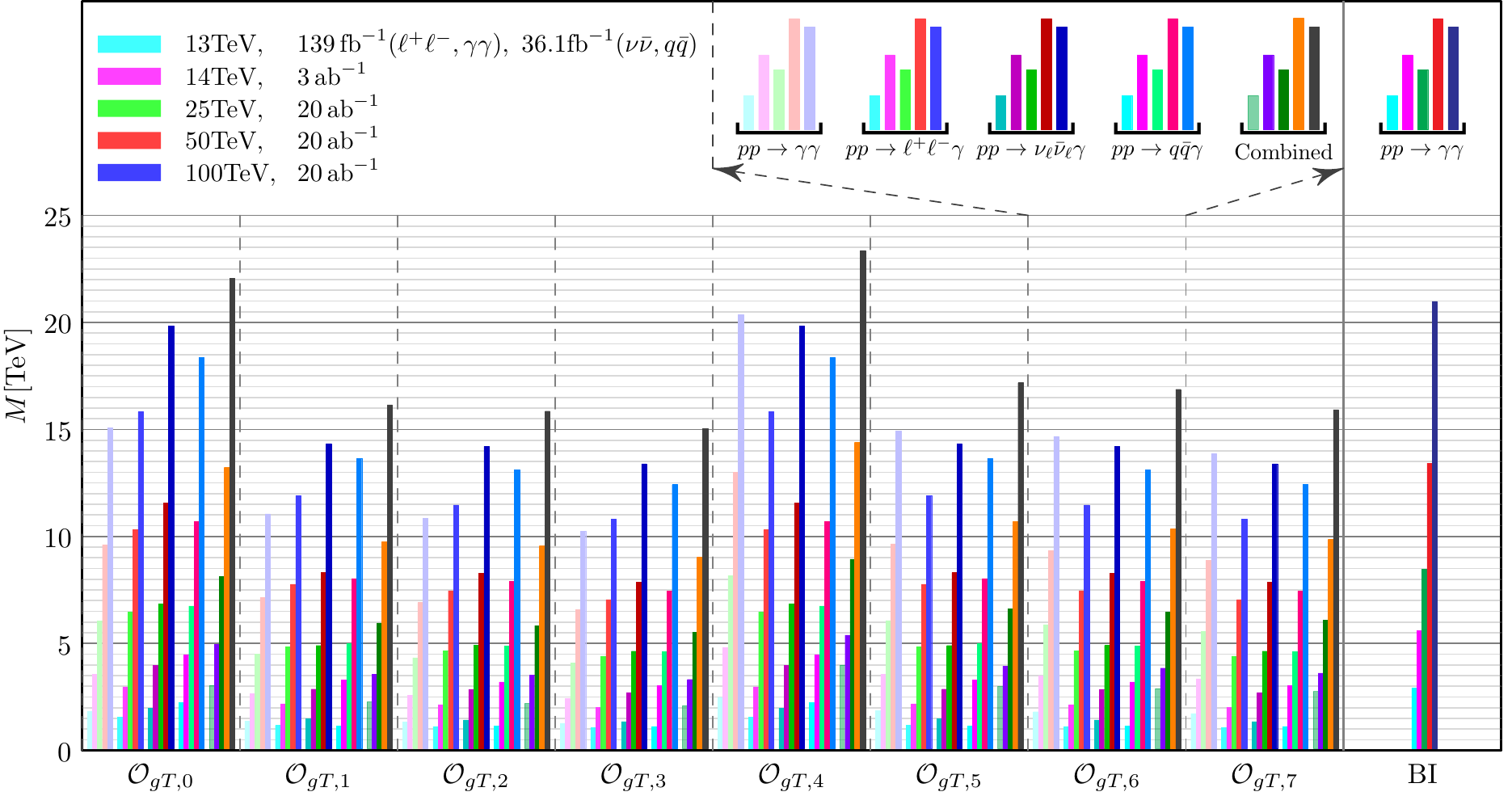}
\caption{\it The 95\% C.L. sensitivities of the various channels at
the LHC, HL-LHC and  proposed future hadron colliders, with cuts 
given in Table~\ref{tab:hptc}.}
\label{fig:combined}
\end{figure}

\gfig{fig:combined}
compares the sensitivities at the LHC and possible future hadron
colliders with centre-of-mass energies of (25, 50, and 100)\,TeV, characteristic of
HE-LHC and FCC-hh/SppC. The sensitivities of the different colliders
are colour-coded with shadings to distinguish the various final states
considered. 
As expected, the higher the collision energy, the greater the sensitivity.
The greatest $Z \gamma$ sensitivity is in general provided by
$Z \to \nu\bar\nu$ decays, while the $\gamma \gamma$
channel provides even greater sensitivities for
$\mathcal O_{gT, (4,5,6,7)}$.

We also show the combined sensitivities of
the $Z \gamma$ and $\gamma \gamma$ channels
\gred{by summing the $\chi^2$ functions in \gfig{fig:chi2AA},
\gfig{fig:ZA:2L:LHC13}, \gfig{fig:ZA:2N:LHC13}, and
\gfig{fig:ZA:2J:LHC13} for the LHC and similarly for the
future higher-energy colliders.}
The lower bound at
95\% C.L. can be significantly enhanced in 100~TeV collisions
to above 20\,TeV for ${\cal O}_{gT, (0,4)}$ and the SM Born-Infeld
extension, an order of magnitude beyond the current LHC limits.

\section{Potential Improvements using the $Z$ Spin and Angular Correlation}
\label{sec:improvements}

As mentioned above, the $pp \rightarrow Z(\ell^+ \ell^-, \bar q q) \gamma$ channels have
three detectable particles in the final state. In these cases, not only the
scattering angle distribution at the intermediate $Z \gamma$ level
that is shown in \gfig{fig:dxsec} can be used for background
discrimination, but also the difference between the $Z \gamma$ scattering plane
and the decay plane of $Z \rightarrow \ell^+ \ell^-, q \bar q$
can provide useful information. Thus the phenomenology of the $Z(\ell^+ \ell^-, \bar q q)$
channels is in principle
much richer than that for $pp \rightarrow \gamma \gamma$ and
$pp \rightarrow Z (\nu \bar \nu) \gamma$, which yield only photons in the
final state. In this section we explore basic features of the $Z$ spin and angular correlations
with a view of potential improvements in the analysis sensitivity.

In order to keep full information on the $Z \rightarrow \ell^+ \ell^-, q \bar q$
decays, we adopt the helicity amplitude formalism \cite{Murayama:1992gi}
to calculate the scattering matrix element for the
$gg \rightarrow Z \gamma$ process:

\begin{subequations}
\begin{align}
\label{eq:proc:prod}
&g( \vec{p}_{1}, \lambda_{1} ) + g( \vec{p}_{2}, \lambda_{2} )
\;\;\; \longrightarrow \;\;\;
Z^{\ast}( \vec{p}_{Z}, \lambda_{Z} ) + \gamma( \vec{p}_{\gamma}, \lambda_{\gamma} ) \, \\[3mm]
\label{eq:proc:decay}
&\hspace{5.6cm}
f(\vec{p}_{f}, \lambda_{f}) + \overline{f}(\vec{p}_{\bar{f}}, \lambda_{\bar{f}})\,.
\begin{picture}(0,0)(111,0)
\put(-10,16){\line(0,-1){13}}
\put(-11.12,0.4){$\longrightarrow$}
\end{picture}
\end{align}
\label{eq:proc}
\end{subequations}
\noindent
Here, $\lambda_{1,2}=\pm1$ and $\lambda_{\gamma}=\pm1$ are the helicities
of the massless gluons and photon, respectively, whereas
in the case of the massive $Z$ boson we have $\lambda_{Z}=0,\,\pm 1$. 
Also, $\lambda_{f}/2=\pm1/2$ denote the helicities of the
final-state fermions from $Z$ decay.
For simplicity, we assume
that the masses of the final-state fermions can be neglected.

The amplitude for the two-step process of $Z \gamma$ production followed
by $Z$ boson decay can be written as
\bee
\label{eq:Amp:total}
\calm 
= 
\cald_{\mu}( Z^{\ast} \to  f + \overline{f})
\calg^{\mu\nu}(Z^{\ast})
\calp_{\nu}( g + g \to Z^{\ast}+ \gamma ),
\ene
where $\calp_{\nu} \equiv \langle Z^{\ast}_{\mu} \gamma|\calo_{i}| g_{1}g_{2}\rangle$
and $\cald_{\mu} \equiv \langle  f\bar{f} |\calo_{SM}| Z^{\ast}_{\nu}\rangle$ are 
the matrix elements for the  production and decay of $Z^{\ast}$, respectively.
The intermediate $Z$-boson propagator $\calg^{\mu\nu}(Z^{\ast})$ can be written as
\bee
\calg^{\mu\nu}(Z^{\ast}) 
\equiv 
\frac{ -\big( g^{\mu\nu}  - p^{\mu}p^{\nu}/m_{Z}^{2} \big) }{ \hat{s} - m_{Z}^{2} + im\varGamma_{Z} }\,.
\ene
Since the decay  width  of the $Z$ boson  is  considerably  smaller  than its  mass,
we use here the narrow-width approximation 
$\big(\hat{s} - m_{Z}^{2} + im\varGamma_{Z}\big)\calg^{\mu\nu}(Z^{\ast}) \approx \sum_{ \lambda_{Z} } \epsilon^\mu_{ \lambda_{Z} }(p) \epsilon^{\nu *}_{ \lambda_{Z} }(p)$.
Then one polarization vector $\epsilon^{\nu *}_{\lambda_Z}$ contracts with $\mathcal P_\nu$
to give the production matrix element while the other polarization vector
$\epsilon^\mu_{\lambda_Z}$ contracts with $\mathcal D_\mu$ to give the decay
matrix element. Consequently, the total helicity amplitude \eqref{eq:Amp:total}
can be decomposed into a sum of products of three scalars:
\bee
\calm 
= 
D_{Z}^{-1}\sum_{ \lambda_{Z} } \calp_{ \lambda_{Z} } \cald_{ \lambda_{Z} } \,,
\label{eq:Mdecomposed}
\ene
where $D_Z \equiv \hat{s} - m_{Z}^{2} + im_{Z}\varGamma_{Z}$. The summation
is over the helicity of the internal $Z$ boson, which does not change during
free propagation. The production and decay matrix elements are
\bea
 \calp_{ \lambda_{Z} } 
\equiv
 \calp( g + g \to Z^{\ast}+ \gamma ) \cdot \epsilon^{\ast}_{ \lambda_{Z} }(p)  \,,
\qquad
 \cald_{ \lambda_{Z} } 
\equiv
 \cald( Z^{\ast} \to   f + \overline{f}\,  ) \cdot \epsilon_{ \lambda_{Z} }(p) \,,
\ena
which can be calculated in the usual way for $gg \rightarrow Z \gamma$
and $Z \rightarrow f \bar f$ separately.

The spin information appears in the absolute square 
of the scattering matrix element,
\bee
  |\mathcal M|^2
=
\big|D_{Z}\big|^{-2}
\sum_{ \lambda_{Z} } \sum_{ \lambda_{Z}^{\prime} }  
\calp_{ \lambda_{Z} }^{ \lambda_{Z}^{\prime} } 
\cald_{ \lambda_{Z} }^{ \lambda_{Z}^{\prime} } \,.
\ene
where
\begin{subequations}
\bea
&&
\calp_{ \lambda_{Z} }^{ \lambda_{Z}^{\prime} } 
\equiv
\calp_{ \lambda_{Z} } \big( \calp_{ \lambda_{Z}^{\prime} } \big)^{\dag}
=
\calp_{\mu}\calp_{\nu}^{\dag} 
\,\epsilon^{\mu\ast}_{ \lambda_{Z} }(p) \, \epsilon^{\nu}_{ \lambda_{Z}^{\prime} }(p)
\equiv
\calp_{\mu\nu}  \,\epsilon^{\mu\ast}_{ \lambda_{Z} }(p) \, \epsilon^{\nu}_{ \lambda_{Z}^{\prime} }(p) \,,
\\[3mm]
&&
\cald_{ \lambda_{Z} }^{ \lambda_{Z}^{\prime} } 
\equiv
\cald_{ \lambda_{Z} } \big( \cald_{ \lambda_{Z}^{\prime} } \big)^{\dag}
=
\cald_{\mu}\cald_{\nu}^{\dag} 
\,\epsilon^{\mu\ast}_{ \lambda_{Z} }(p) \, \epsilon^{\nu}_{ \lambda_{Z}^{\prime} }(p)
\equiv
\cald_{\mu\nu}  \,\epsilon^{\mu}_{ \lambda_{Z} }(p) \, \epsilon^{\nu\ast}_{ \lambda_{Z}^{\prime} }(p) \,,
\ena 
\end{subequations}
The $Z$-boson polarization vectors encode the spin information.

\begin{figure}[t]
\centering
\includegraphics[width=0.68\textwidth]{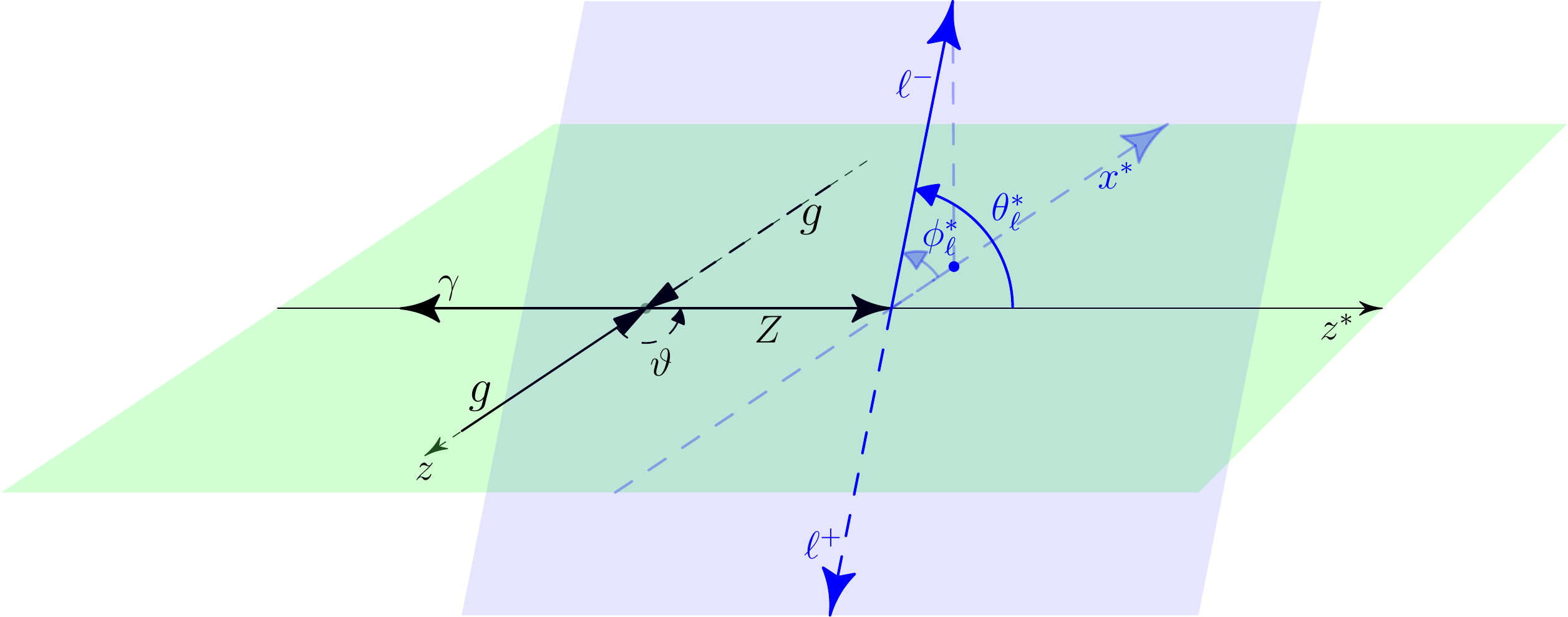}
\caption{\it Kinematic variables for the process $gg \rightarrow Z (f \bar f) \gamma$.
The scattering plane is defined by the incoming gluons and
outgoing $Z/\gamma$ bosons. The polar angle (scattering angle) 
of the $Z$ boson is defined as $\vartheta$, $\vec{z}^{\,\ast}$ is
defined as the $\hat{z}^\ast$-axis in the rest frame of the
$Z$ boson along the direction of the $Z$ boson momentum, and the
$\hat{x}^\ast-\hat{z}^\ast$ plane is spanned by the vector 
$\vec{z}^{\,\ast}$ and the direction of the initial gluon motion.
The polar and azimuthal angles of the leptons in the rest frame
of the $Z$ boson are $\theta_{\ell}^\ast$ and
$\phi_{\ell}^\ast$.
}
\label{fig:kinematics}
\end{figure}

Resolving the decay plane of the channel 
$Z \to \bar \nu \nu$ is not possible, and resolving that of $Z \to \bar q q$
is challenging in practice, particularly when the $Z$ boson is highly boosted.
Therefore here we investigate in detail only the leptonic decay channel.
The $Z$-boson momentum provides a natural definition of the $z$-axis
with respect to which we define the fermion polar angle $\theta^\star_\ell$
in the $Z$ rest frame. Since the detector is azimuthally symmetric, we can
without loss of generality define
the $\hat{x}$-axis to lie in the scattering plane of the $Z$ and $\gamma$ bosons, with
the $Z$ momentum having a positive $\hat{x}$ component.
The azimuthal angle difference between the scattering and decay
planes is $\phi^\star_\ell$, see \gfig{fig:kinematics}.
Then the decay polarization tensor components $\mathcal D^{\lambda'_Z}_{\lambda_Z}$ are
\begin{subequations}
\bea
\cald_{\pm}^{\pm}
&=&
m_{Z}^{2}
\bigg [ \left(  g_{V}^{2}  +  g_{A}^{2}  \right)\,\big( 1 + \cos^{2}\theta^{\star}_{\ell}   \big)
\pm 4 g_{V} g_{A}\cos\theta^{\star}_{\ell}  \bigg] \,,
\\[3mm]
\cald_{0}^{0}
&=&
2m_{Z}^{2} \left( g_{V}^{2}  +  g_{A}^{2}  \right)\, \sin^{2}\theta^{\star}_{\ell} \,,
\\[3mm]
\label{eq:DR:TT}
\cald_{\pm}^{\mp}
&=&
m_{Z}^{2} \left(  g_{V}^{2}  +  g_{A}^{2}   \right)\,
 \sin^{2}\theta^{\star}_{\ell} e^{ \pm i 2 \phi_{\ell}^{\star} } \,,
\\[3mm]
\label{eq:DR:T0}
\cald_{\pm}^{0}
&=&
\frac{\pm1}{\sqrt{2} }
m_{Z}^{2} \Big[  \big( g_{V}^{2}  +  g_{A}^{2}  \big)\sin(2\theta^{\star}_{\ell})  \pm 4  g_{V} g_{A}\sin\theta^{\star}_{\ell}  \Big] e^{ \pm i \phi_{\ell}^{\star} } \,.
\ena
\label{eq:decay}
\end{subequations}
The polarization state of the intermediate $Z$ boson depends in general on the dimension-8
gQGC operators that produces it, carrying information beyond that provided by its momentum.~\footnote{In principle
this also applies to the associated photon,
but it is so energetic that its polarization cannot be
easily extracted.} The decays of the $Z$ boson open a window on the polarization information.
Below we show how to use the fermion polar angle in \gsec{sec:polar} and the
azimuthal angle in \gsec{sec:azimuthal} to distinguish the
gQGC operators. With enough event rate, it is also possible to study
a combined two-dimensional distribution of polar and azimuthal
angles as we discuss in \gsec{sec:combined}.

\subsection{Polarization Effects in the Fermion Polar Angle Distribution}
\label{sec:polar}

The $Z$ polarization tensor $\mathcal P^{\lambda'_Z}_{\lambda_Z}$
is defined in the laboratory frame with the $z$-axis along the beam direction.
Since the scattering is rotationally-invariant with respect to the azimuthal
angle, the components of $\mathcal P^{\lambda'_Z}_{\lambda_Z}$ depend only on the scattering angle $\vartheta$:
\begin{subequations}
\bea
\big(\calp_{0,4}\big)_{\pm}^{\pm}
&=&
\frac{4s_{w}c_{w}}{16^{2} M_{i}^{8} } \frac{\hat{s}^4}{256} (1-x_Z)^2  \Big[ 64 \Big]    \,,
\\[3mm]
\big(\calp_{0,4}\big)_{0}^{0}
&=&
0    \,,
\\[3mm]
\big(\calp_{1,5}\big)_{\pm}^{\pm}
&=&
\frac{4s_{w}c_{w}}{16^{2} M_{i}^{8} } 
\frac{\hat{s}^4}{256} (1-x_Z)^2 \Big[292+ 208 \cos (2 \vartheta )+12 \cos (4 \vartheta )
+ 96 x_Z^2 \sin^4\vartheta \Big]\,,
\\[3mm]
\big(\calp_{1,5}\big)_{0}^{0}
&=&
\frac{4s_{w}c_{w}}{16^{2} M_{i}^{8} } 
\frac{\hat{s}^4}{256} (1-x_Z)^2 
\Big[ 64 \, x_Z \sin^2\vartheta \big(5+ 3\cos (2 \vartheta ) \big) 
\Big]\,,
%
%
\\[3mm]
\big(\calp_{2,6}\big)_{\pm}^{\pm}
&=&
\frac{4s_{w}c_{w}}{16^{2} M_{i}^{8} } 
\frac{\hat{s}^4}{256} (1-x_Z)^2 
\Big[ 163 + 28 \cos (2 \vartheta ) + \cos (4 \vartheta )  + 8 x_Z^{2} \sin^4\vartheta
\Big] \,,
\\[3mm]
\big(\calp_{2,6}\big)_{0}^{0}
&=&
\frac{4s_{w}c_{w}}{16^{2} M_{i}^{8} } 
\frac{\hat{s}^4}{256} (1-x_Z)^2 
\Big[ 16 \, x_Z \sin^2\vartheta \big(3+ \cos (2 \vartheta ) \big) 
\Big] \,,
\\[3mm]
\big(\calp_{3,7}\big)_{\pm}^{\pm}
&=&
\frac{4s_{w}c_{w}}{16^{2} M_{i}^{8} } 
\frac{\hat{s}^4}{256} (1-x_Z)^2 
\Big[ 105 + 20 \cos (2 \vartheta ) + 3\cos (4 \vartheta )  + 24 x_Z^{2} \sin^4\vartheta
\Big] \,,
\\[3mm]
\big(\calp_{3,7}\big)_{0}^{0}
&=&
\frac{4s_{w}c_{w}}{16^{2} M_{i}^{8} } 
\frac{\hat{s}^4}{256} (1-x_Z)^2 
\Big[ 16 \, x_Z \sin^2\vartheta \big(5+ 3\cos (2 \vartheta ) \big) 
\Big]\,,
\ena
\end{subequations}
for the 8 gQGC operators.
We see that the positive ($\lambda_Z = +$) and negative
($\lambda_Z = -$) polarization states are always produced equally.
In other words, there is no net polarization effect. Furthermore,
the production rate of the longitudinal state ($\lambda_Z = 0$)
is suppressed by $x_{Z}$ at large invariant mass $m_{Z\gamma}$,
and therefore can be neglected at high-energy hadron colliders. 
However, in the region $x_{Z} \sim \calo(1)$
it could be useful for distinguishing different Lorentz structures.

\begin{figure}[t]
\begin{center}
\includegraphics[width=0.48\textwidth]{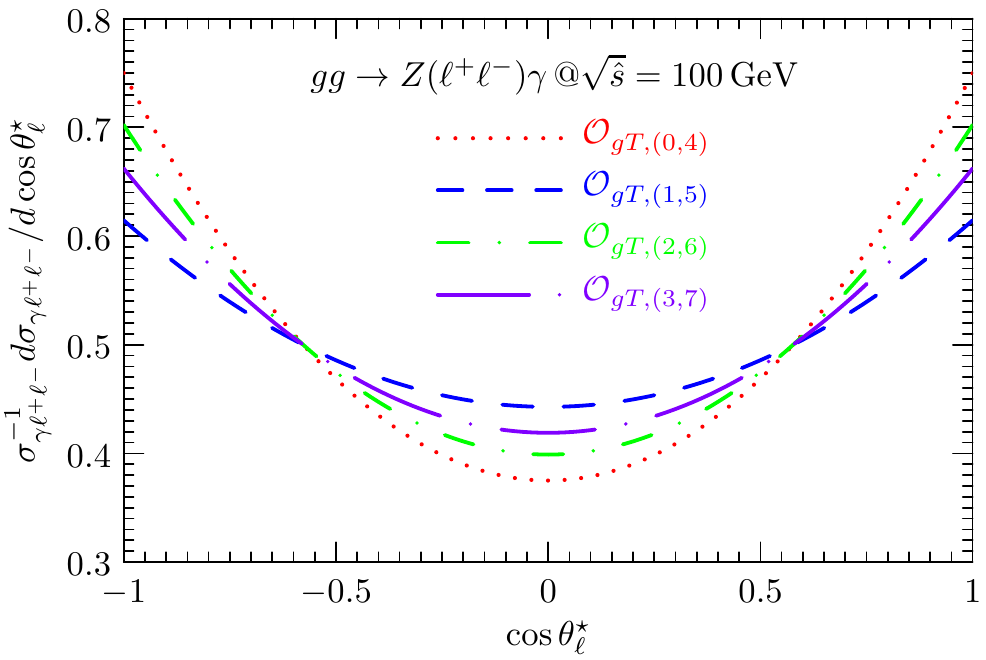}
\quad
\includegraphics[width=0.48\textwidth]{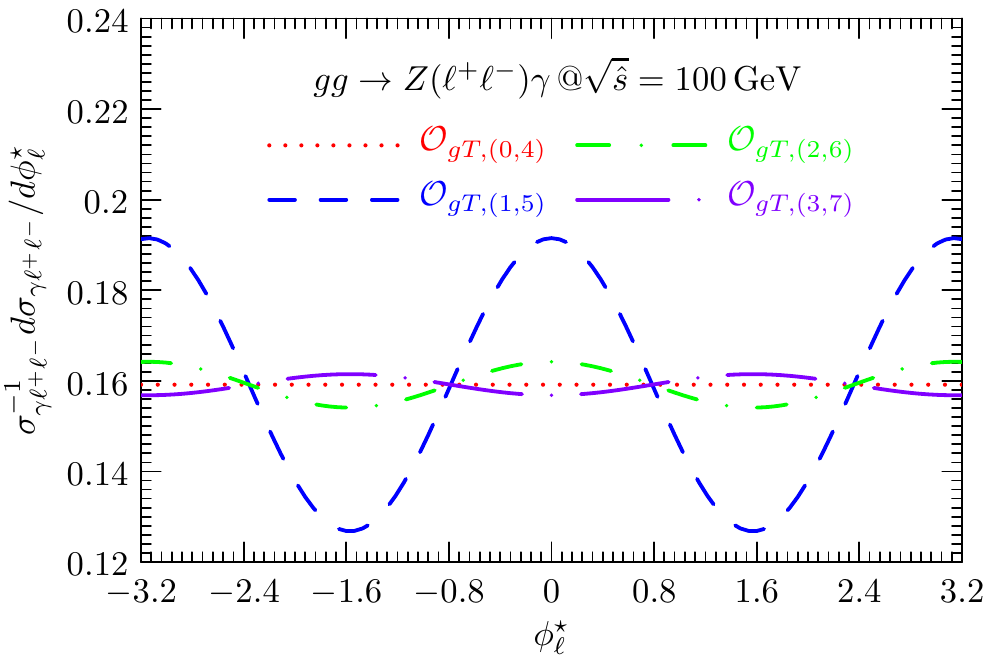}
\caption{\it The normalized differential cross sections in the
polar ($\theta^\star_\ell$) and azimuthal ($\phi^\star_\ell$)
angles of the outgoing lepton
when the gluon-gluon collision energy $\sqrt{\hat{s}} = 100\,\gev$.}
\label{fig:DXS:Pol:Lep}
\end{center}
\end{figure}

In the presence of the longitudinal mode, there is in principle
a polarization effect, though its magnitude
decreases with increasing center-of-mass energy $m_{Z\gamma}$.
The polarization of the $Z$ boson can be measured by studying angular 
distributions of its decay products. The differential cross
sections with respect to the polar angle of the outgoing lepton
are given by
\begin{subequations}
\bea
\frac{d\sigma_{\gamma\ell^{+}\ell^{-}, (0,4) } }{ \sigma_{\gamma\ell^{+}\ell^{-}, 0, 4 } \, d\cos\theta_{\ell}^{\star}} 
&=&
\frac{3}{8} \left( 1 + \cos ^2\theta_{\ell}^{\star} \right) \,,
\\[3mm]
\frac{d\sigma_{\gamma\ell^{+}\ell^{-}, (1,5) } }{ \sigma_{\gamma\ell^{+}\ell^{-}, 1, 5 } \, d\cos\theta_{\ell}^{\star}} 
&=&
\frac{3 \left[26 (1 + \cos^2\theta_{\ell}^{\star})
+ 16 x_Z \sin^2\theta_{\ell}^{\star} +3 x_Z^2 \cos (2 \theta_{\ell}^{\star} )+9 x_Z^2 \right]}{16 \left(13+4 x_Z+3 x_Z^2\right)} \,,
\\[3mm]
\frac{d\sigma_{\gamma\ell^{+}\ell^{-}, (2,6) } }{ \sigma_{\gamma\ell^{+}\ell^{-}, 2, 6 } \, d\cos\theta_{\ell}^{\star}} 
&=&
\frac{3 \left[72(1 + \cos^2\theta_{\ell}^{\star} )
+12 x_Z \sin^2\theta_{\ell}^{\star} +x_Z^2 \cos (2 \theta_{\ell}^{\star} )+3 x_Z^2 \right]}{16 \left(36+3 x_Z+x_Z^2\right)} \,,
\\[3mm]
\frac{d\sigma_{\gamma\ell^{+}\ell^{-}, (3,7) } }{ \sigma_{\gamma\ell^{+}\ell^{-}, 3, 7 } \, d\cos\theta_{\ell}^{\star}} 
&=&
\frac{3 \left[ 46 ( 1+ \cos^2\theta_{\ell}^{\star} )+
16 x_Z \sin^2\theta_{\ell}^{\star} +3 x_Z^2 \cos (2 \theta_{\ell}^{\star} )+9 x_Z^2\right]}{16 \left(23+4 x_Z+3 x_Z^2\right)} \,.
\ena
\end{subequations}
For $\calo_{gT, (0,4)}$, since the $Z$ boson is generated isotropically
as shown in \gfig{fig:dxsec},
the polar angle of the lepton is identical with the distribution in the decay of an
unpolarized $Z$ boson. However, the situations for other operators 
differ because of the contributions of the longitudinal component.
The left panel of \gfig{fig:DXS:Pol:Lep}
shows the normalized differential cross sections
at the gluon-gluon collision energy $\sqrt{\hat{s}} = 100\,\gev$,
where sizeable differences can appear. However, these differences are
proportional to the scale factor $x_{Z}$, and in the limit $x_{Z} \to 0$
all the curves in the left panel of \gfig{fig:DXS:Pol:Lep} converge to forms
$\propto 1+ \cos^2\theta_{\ell}^{\star}$.

\subsection{Spin Correlation in Fermion Azimuthal Angle Distributions}
\label{sec:azimuthal}

Since the intermediate $Z$ is not directly measured,
it is in general produced in an entangled spin state with the
only exception being when produced by the operators $\calo_{gT, (0,4)}$.
Hence different polarizations correlates with each
other and the interference between the helicity states
can induce non-trivial distributions of the fermion
azimuthal angle. To make this transparent, it is necessary
to calculate the relevant off-diagonal matrix elements of
the $gg \to Z \gamma$ production process:
\begin{subequations}
\bea
\big(\calp_{0,4}\big)_{\pm}^{\mp}
&=& 0 \, ,
\\[3mm]
\big(\calp_{1,5}\big)_{\pm}^{\mp}
&=&
\frac{4c_w s_w}{ 16^{2} }  \frac{12}{32} \frac{\hat{s}^4 }{ M_{i}^{8} }
x_Z \left(1-x_Z\right)^2  \big[3 + \cos (2 \vartheta )\big]  \sin^2\vartheta\,,
\\[3mm]
\big(\calp_{2,6}\big)_{\pm}^{\mp}
&=&
\frac{4c_w s_w}{ 16^{2} } \frac{ 1 }{32} \frac{\hat{s}^4 }{ M_{i}^{8} } 
x_Z (1-x_Z)^2 
\big[3 + 2\cos (2 \vartheta )  \big]  \sin^2\vartheta \,,
\\[3mm]
\big(\calp_{3,7}\big)_{\pm}^{\mp} 
&=&
\frac{4c_w s_w}{ 16^{2} } \frac{1}{32} \frac{\hat{s}^4 }{ M_{i}^{8} } 
x_Z (1-x_Z)^2 
\big[1 + 3\cos (2 \vartheta )  \big]  \sin^2\vartheta \,.
\ena
\label{eq:Poff}
\end{subequations}
The related off-diagonal matrix elements of the $Z \to \ell^+\ell^-$
decay process are given in \eqref{eq:DR:TT} and \eqref{eq:DR:T0}.
We see that  the transverse-transverse correlation effect leads to
a non-trivial dependence on the azimuthal angle $\phi_\ell^\star$
of the outgoing lepton. The normalized differential cross sections
with respect to the azimuthal angle of the outgoing lepton for the
different operators are
\begin{subequations}
\bea
\frac{d\sigma_{\gamma\ell^{+}\ell^{-}, 0, 4 } }{ \sigma_{\gamma\ell^{+}\ell^{-}, 0, 4 } \,  d \phi_{\ell}^{\star}}
&=&
\frac{1}{ 2\pi } \,,
\\[3mm]
\frac{d\sigma_{\gamma\ell^{+}\ell^{-}, 1, 5 } }{ \sigma_{\gamma\ell^{+}\ell^{-}, 1, 5 } \,  d \phi_{\ell}^{\star}}
&=&
\frac{26+6 x_Z^2+8 x_Z+9 x_Z \cos (2 \phi_{\ell}^{\star} )}{4 \pi  \left(3 x_Z^2+4 x_Z+13\right)}\,,
\\[3mm]
\frac{d\sigma_{\gamma\ell^{+}\ell^{-}, 2, 6 } }{ \sigma_{\gamma\ell^{+}\ell^{-}, 2, 6 } \, d \phi_{\ell}^{\star}}
&=&
\frac{72 +2 x_Z^2+6 x_Z + 3 x_Z \cos (2 \phi_{\ell}^{\star} )}{4 \pi  \left(x_Z^2+3 x_Z+36\right)} \,,
\\[3mm]
\frac{d\sigma_{\gamma\ell^{+}\ell^{-}, 3, 7 } }{ \sigma_{\gamma\ell^{+}\ell^{-}, 3, 7 } \,  d \phi_{\ell}^{\star}}
&=&
\frac{46 + 6 x_Z^2 + 8 x_Z - x_Z \cos (2 \phi_{\ell}^{\star} ) }{4 \pi  \left(3 x_Z^2+4 x_Z+23\right)} \,.
\ena
\end{subequations}
A non-trivial distribution of the azimuthal angle $\phi_{\ell}^{\ast}$
would indicate the importance of a gQGC operator different from $\mathcal O_{gT, (0,4)}$,
offering in principle the possibility of
distinguishing the gQGC operators. The right panel of \gfig{fig:DXS:Pol:Lep} shows the normalized
differential cross sections at a gluon-gluon collision energy
$\sqrt{\hat{s}} = 100 \, \gev$. We see that, compared to the polarization effect
in the left panel of \gfig{fig:DXS:Pol:Lep}, the correlation effect
is expected to be more sensitive to the Lorentz structures of the
gQGC operators. However, the effect is suppressed
by $x_Z$ at high $M_{Z \gamma}$.

\subsection{Combined Distribution of Scattering and Azimuthal Angles}
\label{sec:combined}

\begin{figure}[t]
\begin{center}
\includegraphics[width=0.48\textwidth]{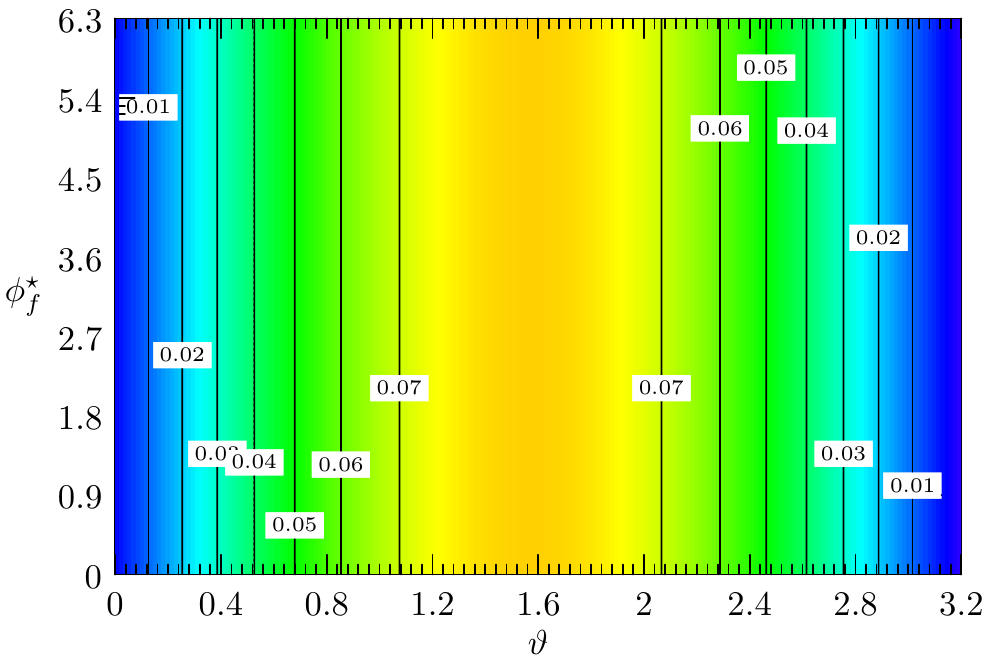}
\quad
\includegraphics[width=0.48\textwidth]{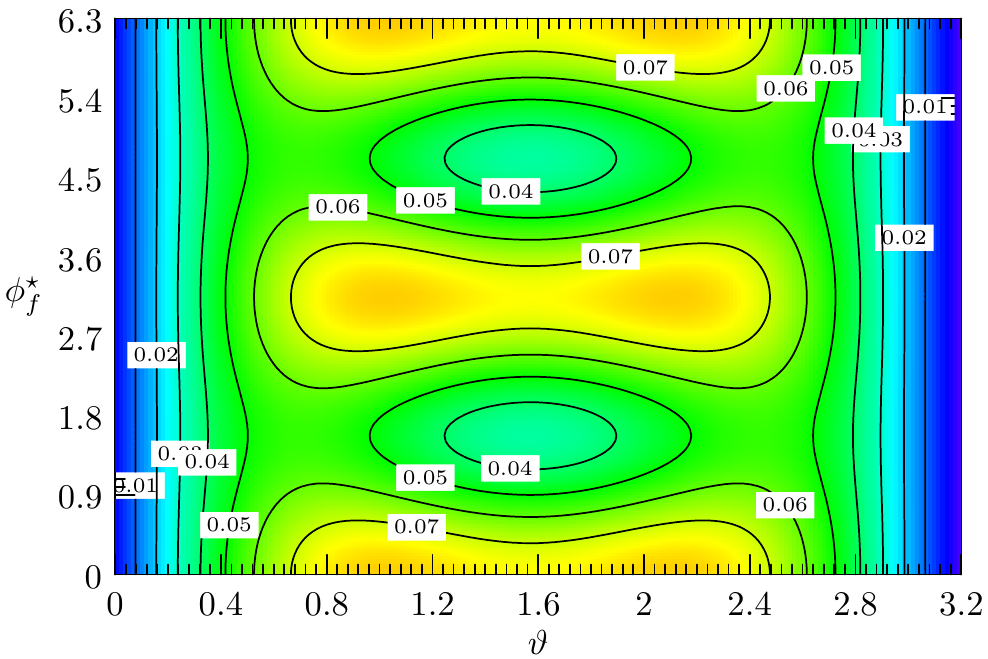}
\\
\includegraphics[width=0.48\textwidth]{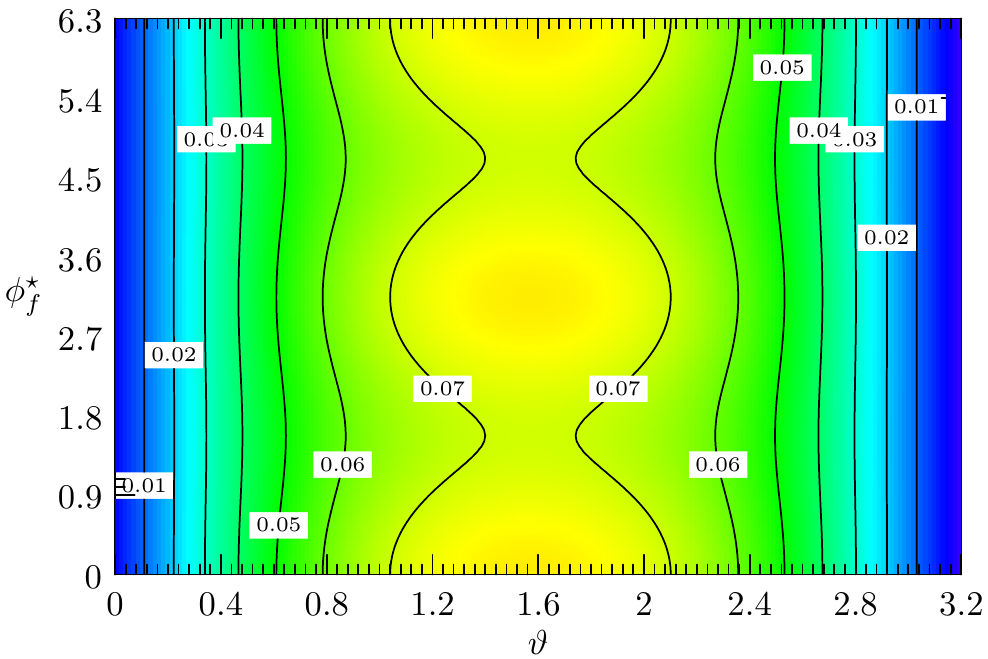}
\quad
\includegraphics[width=0.48\textwidth]{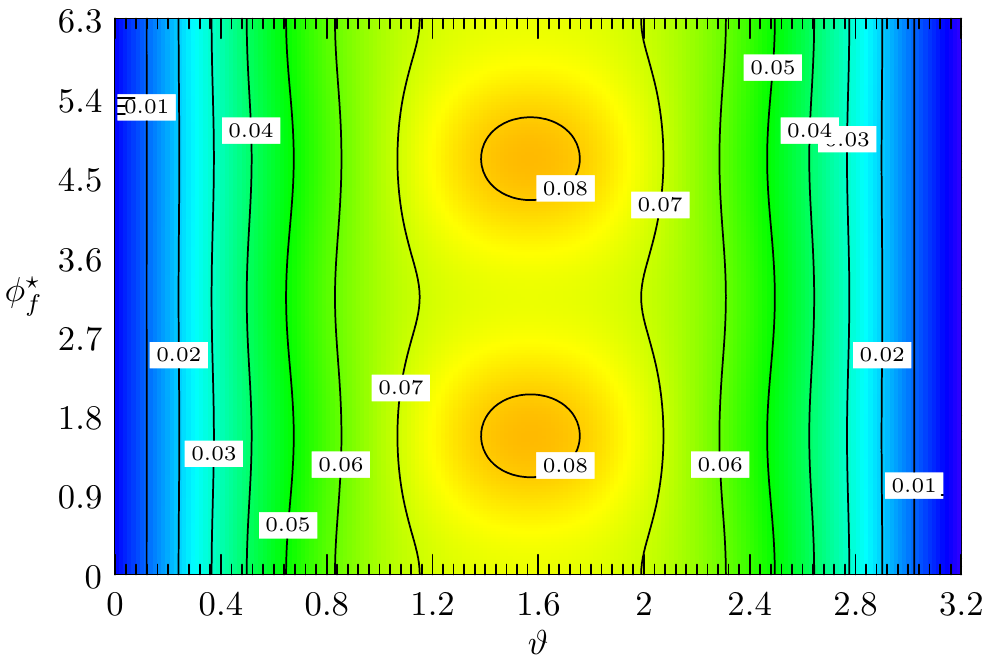}
\label{fig:Cont:TT3}
\caption{\it
Contour plots in the plane of the lepton azimuthal angle
$\phi_{\ell}^{\star}$ and the $Z$ boson polar angle $\vartheta$
for the gQGC operators $\calo_{gT, (0,4)}$ (upper left),
$\calo_{gT,(1,5)}$ (upper right),  $\calo_{gT, (2,6)}$ (lower left),
and $\calo_{gT, (3,7)}$ (lower right). These results are for a
gluon-gluon collision energy $\sqrt{\hat{s}} = 100 \, \gev$.
}
\label{fig:ttc}
\end{center}
\end{figure}

A non-trivial azimuthal angle distribution was also
searched for by ATLAS \cite{ATLAS:2019gey}.
As shown in the previous section, the spin correlation in the
lepton azimuthal angle distribution is due to the off-diagonal
elements \geqn{eq:Poff} in $Z$ production. We observe that these off-diagonal elements have non-trivial
dependence on the $Z$ boson scattering angle $\vartheta$. More
spin information can in principle be extracted from the correlation between
the $Z$ scattering angle $\vartheta$ and the lepton azimuthal
angle $\phi_{\ell}^{\ast}$.
The corresponding normalized double differential cross sections  
for the gQGC operators are
\begin{subequations}
\begin{eqnarray}
\frac{d\sigma_{\gamma\ell^{+}\ell^{-}, 0, 4 } }
{ \sigma_{\gamma\ell^{+}\ell^{-}, 0, 4 } \,   d\cos\vartheta d\phi_{\ell}^{\star} }
& = &
\frac{1}{4\pi} \,,
\nonumber
\\
\frac{d\sigma_{\gamma\ell^{+}\ell^{-}, 1, 5 } }{ \sigma_{\gamma\ell^{+}\ell^{-}, 1, 5 } \,   d\cos\vartheta d\phi_{\ell}^{\star} }
&=&
\frac{ 15 }{ 256 \pi  \left(3 x_Z^2+4 x_Z+13\right) }
\bigg[ 73 + 52 \cos (2 \vartheta )+3 \cos (4 \vartheta ) \,,
\\[3mm]\nonumber
&&
+24 \sin ^4(\vartheta ) x_Z^2+40 \sin ^2(\vartheta ) x_Z+24 \sin ^2(\vartheta ) \cos (2 \vartheta ) x_Z
\\[3mm]
&&
+12 \sin ^2(\vartheta ) \cos (2 \vartheta ) x_Z \cos (2 \phi_{\ell}^{\star} )+36 \sin ^2(\vartheta ) x_Z \cos (2 \phi_{\ell}^{\star} )
\bigg]\,,
\nonumber
\\
\frac{d\sigma_{\gamma\ell^{+}\ell^{-}, 2, 6 } }{ \sigma_{\gamma\ell^{+}\ell^{-}, 2, 6 } \,   d\cos\vartheta d\phi_{\ell}^{\star} }
&=&
\frac{ 15 }{256 \pi  \left(x_Z^2+3 x_Z+36\right) }
\bigg[ 163 + 28 \cos (2 \vartheta )+\cos (4 \vartheta )
\\[3mm]\nonumber
&&
+8 \sin ^4(\vartheta ) x_Z^2+24 \sin ^2(\vartheta ) x_Z+8 \sin ^2(\vartheta ) \cos (2 \vartheta ) x_Z
\\[3mm]
&&
+4 \sin ^2(\vartheta ) \cos (2 \vartheta ) x_Z \cos (2 \phi_{\ell}^{\star} )+12 \sin ^2(\vartheta ) x_Z \cos (2 \phi_{\ell}^{\star} )
\bigg] \,,
\nonumber
\\
\frac{d\sigma_{\gamma\ell^{+}\ell^{-}, 3, 7 } }{ \sigma_{\gamma\ell^{+}\ell^{-}, 3, 7 } \,   d\cos\vartheta d\phi_{\ell}^{\star} }
&=&
\frac{ 15 }{ 256 \pi  \left(3 x_Z^2+4 x_Z+23\right) }
\bigg[ 105 + 20 \cos (2 \vartheta )+3 \cos (4 \vartheta ) 
\\[3mm]\nonumber
&&
+24x_Z \sin^2(\vartheta ) \cos (2 \vartheta ) +24 \sin ^4(\vartheta ) x_Z^2+40 \sin ^2(\vartheta ) x_Z 
\\[3mm]
&&
+12 x_Z \sin ^2(\vartheta ) \cos (2 \vartheta ) \cos (2 \phi_{\ell}^{\star} ) +4 x_Z\sin ^2(\vartheta )  \cos (2 \phi_{\ell}^{\star} )
\bigg] \,.
\end{eqnarray}
\end{subequations}
\gfig{fig:ttc} shows contours of the normalized double-differential cross section
at the gluon-gluon collision energy $\sqrt{\hat{s}} = 100 \, \gev$.
We see that the angles $\vartheta$ and $\phi_{\ell}^{\ast}$ are correlated
very differently for different Lorentz structures, but note that the
transverse-transverse correlation effects also depend on $x_{Z}$.
In particular, for $\calo_{gT, (0,4)}$ and $\calo_{gT, (3,7)}$,
the polar angle $\vartheta$ distributions (at $\sqrt{\hat{s}} = 100 \, \gev$) 
are very similar but the $\phi^\star_\ell$ correlation properties are
very different. Thus the two-dimensional distribution carries additional information beyond that
provided by the one-dimensional distributions.

\section{Conclusions}
\label{sec:conclusion}

The SMEFT is a powerful way of searching for possible indirect effects
of new physics beyond the SM that may appear at energies outside the
direct reaches of active experiments. In particular, the SMEFT
provides a framework for exploring possible anomalous multi-boson couplings
while taking into account all the gauge symmetries of the SM, which is one of the priority
measurements for LHC and possible future colliders. The present and prospective 
experimental sensitivities to anomalous triple gauge couplings (TGCs) and quartic 
gauge couplings (QGCs) have been studied extensively, but there have been fewer
studies of possible quartic couplings between gluons and electroweak gauge bosons,
which are absent in the SM. In the SMEFT, these are first generated by dimension-8
operators, whose study offers an interesting window on BSM physics~\cite{Ellis:2018cos},
complementing the exploration of dimension-6 operators that have been analyzed extensively, see~\cite{Ellis:2020unq, Ethier:2021bye} and references therein.

In this paper we have presented a first analysis of the experimental sensitivities of 
measurements of the $gg \to Z \gamma$ process to the dimension-8 quartic couplings
of pairs of gluons to the $Z$ and photon (gQGCs), including an analysis of present LHC data and
assessments of the prospects at HL-LHC and proposed higher-energy proton-proton colliders. Four 
distinct Lorentz structures of CP-conserving dimension-8 operators contain gQGCs, 
and each may be constructed either with pairs of SU(2) or U(1) gauge field strengths. We have
stressed the differences between the polar angle distributions they generate from
the one generated by the dominant SM background due to $\bar q q \to Z \gamma$, and
analyzed the possible sensitivities of analyses of $Z \to \ell^+ \ell^-, \bar \nu \nu$
and $\bar q q$ final states. The present LHC data on these final states correspond to
integrated luminosities up to 139 fb$^{-1}$ at 13~TeV, and we show that they exclude 
mass scales $\lesssim 2$~TeV in the dimension-8 operator coefficients, with future
colliders offering sensitivities to scales an order of magnitude higher, extending
above 20\,TeV for ${\cal O}_{gT, (0,4)}$.
\gred{The $gg \rightarrow ZZ$ channel has more complex phenomenology
that will be studied in a separate paper.}

We also present an updated analysis of the sensitivities of LHC measurements of the 
$\gamma \gamma$ final state with up to 139 fb$^{-1}$ at 13~TeV and at future colliders.
The present data already constrain the nonlinearity scale of the Born-Infeld extension 
of the SM to be $\gtrsim 5$~TeV, putting pressure on some brane models that address
the electroweak hierarchy problem. A future collider with 100 TeV in the center of mass
could be sensitive to Born-Infeld scales $\gtrsim 20$~TeV.
As we also show, potential analysis improvements might be made possible by exploiting
$Z$ spin measurements and angular correlations using decays into $\ell^+ \ell^-$
and $\bar q q$ final states, which are important at low $Z \gamma$ invariant masses.

Our results provide further illustrations of possibilities for exploring and
constraining possible dimension-8 terms in the SMEFT, in addition to light-by-light
scattering~\cite{Ellis:2017edi, TOTEM:2021kin} and the searches for neutral triple-gauge couplings discussed elsewhere~\cite{Ellis:2019zex, Ellis:2020ljj}.
There is plenty of life in the SMEFT beyond dimension 6.

\section*{Acknowledgements}
The work of J.E. was supported partly by the United Kingdom STFC Grant ST/T000759/1 and partly by the Estonian Research Council via a Mobilitas Pluss grant.
This work was supported by the Double First Class start-up fund
(WF220442604), the Shanghai Pujiang Program (20PJ1407800),
and National Natural Science Foundation of China (No. 12090064).
This work was also supported in part by Chinese Academy of
Sciences Center for Excellence in Particle Physics (CCEPP).
K.M. was supported by the Innovation Capability Support Program of Shaanxi (Program No. 2021KJXX-47), 
as well as the Visitor Program of
Shanghai Jiao Tong University.

\end{document}